\documentclass[aps,prd,preprint,floatfix,superscriptaddress,nofootinbib,longbibliography,a4paper]{revtex4-1}
\pdfoutput=1
\usepackage{color}
\usepackage{graphicx}
\usepackage{textcomp}
\usepackage{subfigure}
\usepackage{feynmp}
\usepackage{amsmath,amssymb,array}
\usepackage{enumerate}
\usepackage{hhline}
\usepackage{hyperref}
\hypersetup{
    colorlinks=true,
    linkcolor=blue,
    filecolor=magenta,      
    urlcolor=blue,}

\newcommand{\beqa}{\begin{eqnarray}}
\newcommand{\eeqa}{\end{eqnarray}}
\newcommand{\be}{\begin{equation}}
\newcommand{\ee}{\end{equation}}
\newcommand{\ba}{\begin{array}} 
\newcommand{\ea}{\end{array}}

\begin{document} 
\vspace*{0.5cm}
\title{Anatomy of scalar mediated proton decays in $SO(10)$ models}
\bigskip
\author{Ketan M. Patel}
\email{kmpatel@prl.res.in}
\affiliation{Theoretical Physics Division, Physical Research Laboratory, Navarangpura, Ahmedabad-380009, India}
\author{Saurabh K. Shukla}
\email{saurabhks@prl.res.in}
\affiliation{Theoretical Physics Division, Physical Research Laboratory, Navarangpura, Ahmedabad-380009, India}
\affiliation{Indian Institute of Technology Gandhinagar, Palaj-382355, India\vspace*{1cm}}

%----------------------------------------------------------
\begin{abstract}
Realistic models based on the renormalizable grand unified theories have varieties of scalars, many of which are capable of mediating baryon ($B$) and lepton ($L$) number non-conserving processes. We identify all such scalar fields residing in ${\bf 10}$, $\overline{\bf 126}$ and  ${\bf 120}$ dimensional irreducible representations of $SO(10)$ which can induce baryon and lepton number violating interactions  through the leading order $d=6$ and $d=7$ operators. Explicitly computing their couplings with the standard model fermions, we derive the effective operators including the possibility of mixing between the scalars stemming from a given representation. We find that such interactions at $d=6$ are mediated by only three sets of scalars: $T(3,1,-1/3)$, ${\cal T} (3,1,-4/3)$ and $\mathbb{T}(3,3,-1/3)$ and their conjugates. In the models with ${\bf 10}$ and $\overline{\bf 126}$, only the first has appropriate couplings to mediate the proton decay. While ${\cal T}$ and $\mathbb{T}$ can induce baryon number violating interactions when ${\bf 120}$ is present, ${\cal T}$ does not contribute to the proton decay at tree level because of its flavour antisymmetric coupling. Three additional colour triplets and their conjugates can mediate nucleon decay via $d=7$ operators which violate also the $B-L$. We give general expressions for partial widths of proton in terms of the fundamental Yukawa couplings and use these results to explicitly compute the proton lifetime and branching ratios for the minimal non-supersymmetric  $SO(10)$ model based on ${\bf 10}$ and $\overline{\bf 126}$ Higgs. We find that the proton preferably decays into $\overline{\nu}\, K^+$ or $\mu^+\, K^0$ and list several distinct features of scalar mediated proton decay. If the latter dominates over the gauge mediated contributions, the proton decay spectrum provides a direct probe to the flavour structure of the underlying grand unified theory.
\end{abstract}
%----------------------------------------------------------

\maketitle

\section{Background}
\label{sec:intro}
Unification of the Standard Model (SM) gauge symmetries in simple gauge groups \cite{Fritzsch:1974nn,Georgi:1974sy,Gell-Mann:1979vob} often implies the existence of a rich spectrum of the scalars beyond the electroweak doublet Higgs. The latter is just one of the many submultiplets of an irreducible representation (irrep) of the underlying unified symmetry group. The dimensions of these representations are typically larger than the ones which unify, partially or completely, the SM quarks and leptons of a given generation. This is particularly the case in renormalizable models in which more than one type of Yukawa interactions are needed to reproduce the realistic flavour spectrum including neutrino masses and mixing parameters \cite{Dimopoulos:1991yz,Babu:1992ia,Clark:1982ai,Aulakh:1982sw,Aulakh:2003kg}. For example, in renormalizable models based on $SO(10)$ Grand Unified Theory (GUT), one needs at least two of the three, namely ${\bf 10}$, $\overline{\bf 126}$ and ${\bf 120}$, irreps for this purpose \cite{Bajc:2005zf,Joshipura:2011nn}. They respectively contain 4, 22 and 24 numbers of SM scalar multiplets with diverse colour and electroweak charges, see Table \ref{tab:scalars}. Their non-trivial charges under the SM gauge symmetries and couplings with the SM quarks and leptons make them interesting from the phenomenological considerations. Potential new physics effects of many of these scalars have been explored in the context of the recent flavour anomalies (see for example, \cite{Bordone:2016gaq,Belanger:2021smw,Perez:2021ddi,Sahoo:2021vug}), muon $g-2$ anomaly \cite{Bauer:2015knc,Chen:2017hir,Dorsner:2019itg}, some events at the direct search experiments \cite{Dorsner:2009mq,Patel:2011eh,Dorsner:2016wpm}, GUT scale baryogenesis \cite{Babu:2012iv,Babu:2012vb} and precision unification of gauge couplings in the absence of low energy supersymmetry \cite{FileviezPerez:2008afb,Dorsner:2009mq,Patel:2011eh,Babu:2012iv,Babu:2012vb}. All these require at least a few scalars at the sub GUT scale.

Some of the scalars with non-zero $B-L$ charges are capable of mediating baryon number ($B$) and lepton number  ($L$) violating decays of baryons  \cite{Pati:1973uk,PhysRevLett.43.1566,PhysRevLett.43.1571} (see also \cite{Langacker:1980js,Nath:2006ut,Babu:2013jba,Raby:2017ucc,Fukuyama:2004ps} for reviews) depending on their couplings with the SM quarks and leptons. In typical bottom-up approaches, the freedom to choose their couplings is often exploited to avoid dangerously fast proton decays. However, if these scalars are part of GUT multiplets which also contains the SM Higgs like fields, their couplings are often the same couplings that determine the low energy quark and lepton mass spectrum.  Hence, there may exist strong constraints on the masses of these scalar fields in these models from the stability of the proton. Such constraints may then forbid them from being light from the perspective of top-down approaches. Therefore, it is important to derive explicitly the exact nature of the couplings of various scalars in the underlying GUT framework and their precise predictions for the nucleon decay spectrum. Such an analysis has been carried out earlier for $SU(5)$ and flipped $SU(5)$ GUTs in \cite{Golowich:1981sb,Dorsner:2012nq}. We perform a comprehensive study for a class of renormalizable $SO(10)$ GUTs which has a relatively richer spectrum of scalars and more constrained coupling structure for Yukawa interactions than those in $SU(5)$ GUTs because of a complete unification of quarks and leptons.

Scalar induced contribution to the proton decay has received less attention in comparison to the one mediated by vector bosons in GUTs. The latter has been extensively studied including the flavour effects \cite{Dorsner:2004jj,FileviezPerez:2004hn,Dorsner:2004xa,Kolesova:2016ibq}. Firstly, this is because the vector boson induced contributions dominate over the scalar mediated ones if the mass scale of both the mediators is of similar magnitude as the latter are suppressed by the first generation Yukawa couplings. The ratio of scalar to vector induced contributions in the proton decay width is naively given by
\be \label{scal_vec} 
\frac{y_u^4/M_S^4}{g^4/M_V^4} \sim \left({\cal O}(10^{-4})\, \frac{M_V}{M_S} \right)^4\,,\ee
where $g$ is unified gauge coupling, $y_u$ is up-type quark Yukawa coupling and $M_{V,S}$ are masses of vector and scalar mediators, respectively. Therefore, scalar contributions becomes significant only if $M_S \lesssim {\cal O} (10^{-4})\, M_V$. Secondly, the scalar mediated contributions depend on the Yukawa sector of the theory and are model-dependent to a large extent\footnote{The vector boson induced contributions also depend on the Yukawa interactions through unitary rotations which relate physical basis with the field basis. This dependency, however, is indirect and relatively mild in comparison to the scalar mediated contributions.}. Many of the important aspects of the scalar induced nucleon decays depend on the nature of the mediator and its exact couplings with the SM quarks and leptons. These details are not completely captured in the generic model-independent effective theory based analysis \cite{PhysRevLett.43.1566,PhysRevLett.43.1571}.

This work is aimed to provide a complete classification of the scalar induced nucleon decays in the models based on renormalizable $SO(10)$ GUTs. Identifying various scalar fields residing in ${\bf 10}$, $\overline{\bf 126}$ and ${\bf 120}$ which  carry $B-L$ current, we determine their explicit couplings with the SM quarks and leptons with appropriate Clebsch-Gordan coefficients. Interactions that can give rise to nucleon decays at tree level  through effective dimension-6 ($d=6$) and dimension-7 ($d=7$) are then listed out and used to determine explicitly the proton decay widths in terms of the fundamental Yukawa couplings. In the predictive versions of these GUTs, the latter can be determined, partially or fully, from the observed masses and mixing parameters of the quarks and leptons. This enables computation of the proton decay widths with less ambiguity in comparison to the effective theory based estimations. We demonstrate this by estimating proton decays for the minimal non-supersymmetric $SO(10)$ GUT which uses ${\bf 10}_H$ and $\overline{\bf 126}_H$ to account for realistic fermion spectrum. The proton decay pattern is found significantly different from the one induced by vector bosons in this case.

We give a complete spectrum  of various scalar sub-multiplets which may arise in renormalizable $SO(10)$ models in the next section and compute their couplings relevant for nucleon decays. In sections \ref{sec:operators_d6} and \ref{sec:operators_d7}, we derive effective operators after integrating out the relevant scalars and give the explicit expressions of proton decay widths in terms of these effective couplings in section \ref{sec:decay_widths}. These results are used to compute the scalar mediated proton decay spectrum in a minimal model based on two GUT representations in the section \ref{sec:results_model}. Finally, we summarize in section \ref{sec:summary}.

\section{Scalar spectrum and couplings}
\label{sec:couplings}
In the renormalizable versions of $SO(10)$ gauge theory, the Yukawa sector, in general, can consist of scalars which are ${\bf 10}$, $\overline{\bf 126}$ and ${\bf 120}$ dimensional irreps of the underlying gauge group. These irreps, when decomposed under the SM gauge group, contain several multiplets charged under $B-L$ which can give rise to baryon and/or lepton number violating interactions. The SM and $B-L$ charges of these fields and their multiplicity in ${\bf 10}_H$, $\overline{\bf 126}_H$ and ${\bf 120}_H$ are listed in Table \ref{tab:scalars}. We use convention in which,
\be \label{BL}
B-L = \frac{4}{5}Y-\frac{1}{5}X\,,\ee
where $X$ and $Y$ are the charges under $U(1)_X$ and $U(1)_Y$ subgroups of $SO(10)$, respectively \cite{Buchmuller:2019ipg}. The hypercharge generator is normalized in such a way that the electric charge $Q = T_3 + Y$.
%%%%%%%%%%%%%
\begin{table}[t]
\begin{center}
\begin{tabular}{cccccc} 
\hline
\hline
~~SM charges~~&~~Notation~~&~~$B-L$~~&~~${\bf 10}_H$~~&~~$\overline{\bf 126}_H$~~&~~${\bf 120}_H$~~\\
 \hline
$\left(1,1,0\right)$ &$\sigma$ &$-2$ &$0$ &$1$ &$0$ \\
$\left(1,1,1\right)$ & $s$ &$2$ &$0$ &0 &$1$\\
$\left(1,1,-1\right)$ & $\overline{s}$ &$-2$  &0 &1 &1\\
$\left(1,1,-2\right)$   &$X$ &$-2$ &0 &1 &0\\
$\left(1,2,-\frac{1}{2}\right)$ & $\overline{D}_a$ &0 &1 &1 &2\\
$\left(1,2,\frac{1}{2}\right)$ & $D^a$ &0 &1 &1 &2\\
$\left(1,3,1\right)$ & $t^{ab}$ &2 &0 &1 &0\\
$\left(3,1,-\frac{1}{3}\right)$ & $T^{\alpha}$ &$-\frac{2}{3}$ &1 &2 &2\\
$\left(\overline{3},1,\frac{1}{3}\right)$ &$\overline{T}_\alpha$ & $\frac{2}{3}$&1 &1 &2\\
$\left(3,1,\frac{2}{3}\right)$ & $\Theta_{\alpha\beta}$ & $-\frac{2}{3}$ &0 &1 &1\\
$\left(\overline{3},1,-\frac{2}{3}\right)$ &$\overline{\Theta}^{\alpha\beta}$ &$\frac{2}{3}$ &0 &0 &1\\
$\left(3,1,-\frac{4}{3}\right)$ & $\mathcal{T}^\alpha$ & $-\frac{2}{3}$ &0 &1 &1\\
$\left(\overline{3},1,\frac{4}{3}\right)$&$\overline{\mathcal{T}}_\alpha$ &$\frac{2}{3}$ &0 &0 &1\\
$\left(3,2,\frac{1}{6}\right)$ & $\Delta^{\alpha a}$ & $\frac{4}{3}$ &0 &1 &1\\
$\left(\overline{3},2,-\frac{1}{6}\right)$& $\overline{\Delta}_{\alpha a}$&$-\frac{4}{3}$ & 0&1 &1\\
$\left(3,2,\frac{7}{6}\right)$ & $\Omega^a_{\alpha \beta}$ & $\frac{4}{3}$ &0 &1 &1\\
$\left(\overline{3},2,-\frac{7}{6}\right)$&$\overline{\Omega}_a^{\alpha \beta}$ & $-\frac{4}{3}$&0 &1 & 1\\
$\left(3,3,-\frac{1}{3}\right)$ & $\mathbb{ T}{^{a\alpha}_{b}}$ & $-\frac{2}{3}$ &0 &0 &1\\
$\left(\overline{3},3,\frac{1}{3}\right)$&$\overline{\mathbb{T}}{^a_{b\alpha}}$ &$\frac{2}{3}$ &0 &1 &1\\
$\left(6,1,\frac{1}{3}\right)$ & $S{^{\alpha}_{\beta\gamma}}$ & $\frac{2}{3}$ &0 &1&1\\
$\left(\overline{6},1,-\frac{1}{3}\right)$&$\overline{S}{^{\beta\gamma}_{\alpha}}$ &$-\frac{2}{3}$ &0 &0 &1 \\
$\left(6,1,-\frac{2}{3}\right)$ & $ \Sigma^{\alpha\beta}$ & $\frac{2}{3}$ &0 &1 &0 \\
$\left(6,1,\frac{4}{3}\right)$ & ${\cal S}^{\alpha}_{\beta\gamma}$ & $\frac{2}{3}$ & 0 & 1 & 0 \\
$\left(\overline{6},3,-\frac{1}{3}\right)$ &$\mathbb{S}^{\alpha\beta a}_{\gamma b}$ & $-\frac{2}{3}$ &0 &1 &0\\
$\left(8,2,\frac{1}{2}\right)$ & $O^{\alpha a}_{\beta}$ &0 &0 &1 &1\\
$\left(8,2,-\frac{1}{2}\right)$ & $\overline{O}_{\alpha a}^{\beta}$ &0 &0 &1 &1\\
\hline
\end{tabular}
\end{center}
\caption{Classification of different scalar fields residing in ${\bf 10}_H$, $\overline{\bf 126}_H$ and ${\bf 120}_H$ with their charges under the SM gauge group ($SU(3)_C$, $SU(2)_L$, $U(1)_Y$), $B-L$ charge and multiplicities in the given GUT representation.}
\label{tab:scalars}
\end{table}
%%%%%%%%%%%%%%%%%%%%

To derive couplings of these scalars with the SM quarks and leptons, we decompose each of the $SO(10)$ invariant Yukawa terms into the ones invariant under the SM gauge symmetry. This is done by first decomposing the underlying $SO(10)$ interactions into $SU(5)$ and then the latter is further decomposed in terms of the interactions involving the SM fermions and appropriately normalised scalars. In the following, we use $i,j,k,...$ to represent $SU(5)$ tensors, $\alpha,\beta,\gamma,...$ and $a,b,c,...$ to represent $SU(3)$ and $SU(2)$ tensors, respectively. The three generations of fermions are denoted by $A,B=1,2,3$. The irreps of $SO(10)$ are distinguished from those of $SU(5)$ by indicating the former in bold.

\subsection{${\bf 16}_F$-${\bf 16}_F$-${\bf 10}_H$ couplings}
The interactions of ${\bf 16}$-plet fermions with ${\bf 10}_H$ is parametrized by
\be \label{16-16-10}
-{\cal L}^{\bf 10}_Y = H_{AB}\,{\bf 16}^T_A\,C^{-1}\,{\bf 16}_B\,{\bf 10}_H\,+\,{\rm h.c.}\,, \ee
where $H_{AB} = H_{BA}$. $C$ is the Lorentz charge conjugation matrix. For brevity, we have suppressed the equivalent matrix in the gauge space while writing the above term. For complex scalar ${\bf 10}_H$, one can also have an additional gauge invariant Yukawa term with ${\bf 10}_H^\dagger$. However, to derive the couplings of scalars with SM quarks and leptons, it is sufficient to use one of these. The $SO(10)$ invariant terms can be translated into the $SU(5)$ language following a procedure discussed in \cite{PhysRevD.21.1062,Nath:2001uw,Syed:2005gd}. Under the $SU(5)$ decomposition, Eq. (\ref{16-16-10}) becomes
\be \label{16-16-10_su5}
-{\cal L}^{\bf 10}_Y = i 2\sqrt{2}\, H_{AB}\,\left({10^{ij}_A}^T\, C^{-1}\,\overline{5}_{i B}\, \overline{5}_j\, +\, \frac{1}{8} \epsilon_{ijklm}\, {10^{ij}}^T_A\, C^{-1}\,10^{kl}_B\, 5^m\, -\, 1_A^T\, C^{-1}\,\overline{5}_{iB}\, 5^i \right)\,+\,{\rm h.c.}\,. \ee
All the $SU(5)$ representations appearing above are defined in such a way that their kinetic terms are normalized.

The $SU(5)$ fields are further decomposed in the SM fields as the following. The fermions residing in $10$, $\overline{5}$ and $1$ dimensional irreps can be identified with the SM fermions as
\be \label{10ferm_su5}
10^{\alpha \beta} \equiv \epsilon^{\alpha \beta \gamma}\,u^C_\gamma\,,~~10^{\alpha a} \equiv q^{\alpha a}\,,~~10^{ab} \equiv \epsilon^{ab}\,e^C\,,\ee
\be \label{5ferm_su5}
\overline{5}_\alpha \equiv d^C_\alpha\,,~~\overline{5}_{a} \equiv \epsilon_{ab}\,l^b\,,~~1 \equiv \nu^C\,,\ee
where all the fermion fields are left-handed Weyl fermions. Similarly, the scalars identified in $5$- and $\overline{5}$-plet can be written as 
\beqa \label{5dec}
5^\alpha \equiv T^\alpha\,,~~5^a \equiv D^a\,,~~\overline{5}_\alpha \equiv \overline{T}_\alpha\,,~~\overline{5}_a \equiv \overline{D}_a\,. \eeqa
such that their kinetic terms are also in the canonically normalized form. Substitution of Eqs. (\ref{10ferm_su5},\ref{5ferm_su5},\ref{5dec}) in Eq. (\ref{16-16-10_su5}) leads to the following terms involving $B-L$ charged scalar fields. 
\beqa \label{10/5}
-{\cal L}^{{\bf 10}/5}_Y & \supset & i 2\sqrt{2}\, H_{AB}\, \left( u^{C\,T}_{\gamma A}\, C^{-1}\,e^C_B  -\frac{1}{2} \epsilon_{\alpha \beta \gamma}\, \epsilon_{ab}\, q^{\alpha a\, T}_A\, C^{-1}\,q^{\beta b}_B \right)T^\gamma\,+\,{\rm h.c.}\,, \nonumber \\
-{\cal L}^{{\bf 10}/\overline{5}}_Y & \supset & i 2\sqrt{2}\, H_{AB}\, \left(\epsilon^{\alpha \beta \gamma}\, u^{C\, T}_{\alpha A}\,C^{-1}\,d^C_{\beta B} - \epsilon_{ab}\, q^{\gamma a\,T}_A\,C^{-1}\,l^b_B\right) \overline{T}_\gamma\,+\,{\rm h.c.}\,. \eeqa
Consequently, both $T$ and $\overline{T}$ have diquark and leptoquark couplings and each can mediate $B$ and $L$ violating decays of baryons.

\subsection{${\bf 16}_F$-${\bf 16}_F$-$\overline{\bf 126}_H$ couplings}
The $\overline{\bf 126}_H$ has a very rich scalar spectrum. The Yukawa Lagrangian involving this field is written as
\be \label{16-16-126}
-{\cal L}^{\overline{\bf 126}}_Y = F_{AB}\,{\bf 16}^T_A\,C^{-1}\,{\bf 16}_B\,\overline{\bf 126}_H\,+\,{\rm h.c.}\,, \ee
where $F_{AB} = F_{BA}$. In the $SU(5)$ notations, the above Lagrangian can be written as \cite{Nath:2001uw}
\beqa \label{16_126_su5}
-{\cal L}^{\overline{\bf 126}}_Y &=& i \frac{2}{\sqrt{15}}\, F_{AB}\,\Big[-1{^T_A}\, C^{-1}\, 1_B\, 1_0 + 1{^T_A}\,C^{-1}\, 10_B^{ij}\, \overline{10}_{ij}-\overline{5}{_{i A} ^T}\, C^{-1}\, \overline{5}_{j B}\, 15^{ij}\, \Big. \nonumber \\
&-& \sqrt{\frac{3}{2}}\,\left(1{^T_A}\, C^{-1}\, \overline{5}_{i B}\, 5^{i}\, + \frac{1}{24}\epsilon_{ijklm}\, 10^{ij\,T}_A\,C^{-1}\, 10^{kl}_B\,5^m \right)\,\nonumber \\
& + & \Big. 10^{ij\,T}_A\, C^{-1}\,  \overline{5}_{k B}\, \overline{45}^k_{ij} - \frac{1}{4\sqrt{3}}\,\epsilon_{ijklm}\, 10^{ij\,T}_A\,C^{-1}\, 10^{rs}_B\,50^{klm}_{rs} \Big]\,+\,{\rm h.c.}\,.\eeqa
The first term above does not involve the SM fermions. Similarly, the second term gives rise to lepto-quark coupling with only the SM singlet leptons which are typically heavier than the nucleon. Hence, both these terms are not relevant for the proton decay.

The $\overline{\bf 126}_H$ contains a $5$-plet of $SU(5)$ whose decomposition is given earlier in Eq. (\ref{5dec}). Using this and the second line in Eq. (\ref{16_126_su5}), we get
\beqa \label{126/5}
-{\cal L}^{\overline{\bf 126}/5}_Y & \supset & i \sqrt{\frac{2}{5}}\, F_{AB}\, \left(- \frac{1}{3} u{^{CT}_{\gamma A}}\, C^{-1}\, e^C_B +\frac{1}{6} \epsilon_{\alpha \beta \gamma}\, \epsilon_{ab}\, q^{\alpha a T}_A\, C^{-1}\, q^{\beta b}_B \right)T_1^\gamma\,+\,{\rm h.c.}\,. \eeqa
Since $\overline{\bf 126}_H$ has more than one color triplet fields, we distinguish them by assigning a subscript. Decomposition of $15$-plet into the normalized irreps of $SU(5)$ is given by
\beqa \label{15dec}
15^{\alpha \beta} = \Sigma^{\alpha \beta}\,,~~15^{\alpha a} = \frac{1}{\sqrt{2}} \Delta^{\alpha a}\,,~~15^{ab} = t^{ab}\,.\eeqa
$\Sigma^{\alpha \beta}$ ($t^{ab}$) is color sextet (weak triplet) and has only diquark (dilepton) couplings. Therefore, they do not contribute in the proton decay. The color triplet weak doublet field $\Delta^{\alpha a}$ has  $Y=1/6$ and its couplings with the SM fermions are obtained as
\beqa \label{126/15}
-{\cal L}^{\overline{\bf 126}/15}_Y & \supset & i 2\sqrt{\frac{2}{15}}\, F_{AB}\, \epsilon_{ab}\, d{^{CT}_{\alpha A}}\, C^{-1}\, l^a_B\, \Delta^{\alpha b}\,+\,{\rm h.c.}\,. \eeqa
$\Delta^{\alpha b}$ has only the lepto-quark coupling and hence it cannot induce proton decay by itself.

Next, we consider $\overline{45}$-plet which decomposes into various irreps of $SU(5)$ as:
\beqa \label{45dec}
\overline{45}^{\gamma}_{\alpha \beta} &\equiv& \frac{1}{\sqrt{2}} S^{\gamma}_{\alpha \beta} + \frac{1}{2 \sqrt{2}} \left( \delta^\gamma_{\alpha} \overline{T}_\beta - \delta^\gamma_\beta \overline{T}_\alpha \right)\,,~~\overline{45}^a_{\alpha \beta}  \equiv  \frac{1}{\sqrt{2}} \Omega^a_{\alpha \beta}\,\,, \nonumber \\
\overline{45}^\beta_{\alpha a} &\equiv& \frac{1}{\sqrt{2}} \overline{O}^\beta_{\alpha a} + \frac{1}{2\sqrt{6}} \delta^\beta_\alpha \overline{D}_a\,,~~\overline{45}^\beta_{ab} \equiv \frac{1}{\sqrt{2}} \epsilon_{ab} {\cal T}^\beta\,,\nonumber \\
\overline{45}^a_{b \alpha} & \equiv & \frac{1}{\sqrt{2}} \overline{\mathbb{T}}^a_{b \alpha} - \frac{1}{2 \sqrt{2}} \delta^a_b \overline{T}_\alpha\,,~~ \overline{45}^{c}_{ab} \equiv -\frac{\sqrt{3}}{2\sqrt{2}} \left( \delta^c_a \overline{D}_b - \delta^c_b \overline{D}_a\right)\,. \eeqa
The SM and $B-L$ charges of these fields are listed in Table \ref{tab:scalars}. All the fields other than $\overline{O}^\beta_{\alpha a}$ and $\overline{D}_a$ carry non-trivial $B-L$ charges. However, $S_{\alpha \beta}^\gamma$ couples to the quarks only while $\Omega{^a_{\alpha\beta}} = (3,2,7/6)$ has only lepto-quark vertex. The couplings of the remaining multiplets with the SM quarks and leptons are obtained as
\beqa \label{126/45}
-{\cal L}^{\overline{\bf 126}/\overline{45}}_Y & \supset & i \sqrt{\frac{2}{15}}\, F_{AB}\, \left[\left(\epsilon^{\alpha \beta \gamma}\, u^{CT}_{\alpha A}\,C^{-1}\,\,d^C_{\beta B} + \epsilon_{ab}\, q^{\gamma a T}_A\,C^{-1}\,l^b_B\right) \overline{T}_\gamma\, \right. \nonumber \\ 
& + & \left. 2\, e^{CT}_A\,C^{-1}\, d^C_{\alpha B}\, {\cal T}^\alpha\,+\, 2\, \epsilon_{bc}\,q^{\alpha a T}_A\,C^{-1}\,l^b_B\,\overline{\mathbb{T}}^c_{a \alpha} \right]\,+{\rm h.c.}\,. \eeqa
Note that the chromo-weak triplet $\mathbb{T}^a_{b \alpha}$ and $Y=-4/3$ color triplet ${\cal T}^\alpha$ also have only lepto-quark coupling at this stage. This, along with quark-quark couplings arising from ${\bf 120}_H$, can induce the nucleon decay if there exists a mixing between these scalars in the underlying model.

Finally, the coupling with $50$-plet of $SU(5)$ can be computed from its following decomposition:
\beqa \label{50dec}
50^{\alpha \beta \gamma}_{\sigma \rho} & \equiv & \frac{1}{6} \left[ \left(\delta^\alpha_\sigma \delta^\beta_\rho -  \delta^\alpha_\rho \delta^\beta_\sigma \right) T^\gamma +  \left(\delta^\alpha_\rho \delta^\gamma_\sigma -  \delta^\alpha_\sigma \delta^\gamma_\rho \right) T^\beta  +  \left(\delta^\beta_\sigma \delta^\gamma_\rho -  \delta^\beta_\rho \delta^\gamma_\sigma \right) T^\alpha \right]\,,\nonumber \\
50^{\alpha \beta \gamma}_{\sigma a} & \equiv & \frac{1}{2\sqrt{3}} \left( \delta^\alpha_\sigma \overline{\Omega}^{\beta \gamma}_a + \delta^\beta_\sigma \overline{\Omega}^{\gamma \alpha}_a + \delta^\gamma_\sigma \overline{\Omega}^{\alpha \beta}_a\right)\,, \nonumber \\
50^{\alpha \beta a}_{\sigma \rho} & \equiv & \frac{1}{2\sqrt{6}} \left(\delta^\alpha_\sigma O^{\beta a}_\rho - \delta^\alpha_\rho O^{\beta a}_\sigma +\delta^\beta_\rho O^{\alpha a}_\sigma - \delta^\beta_\sigma O^{\alpha a}_\rho \right)\,,\nonumber\\
50^{\alpha \beta \gamma}_{ab} & \equiv & \frac{1}{2 \sqrt{3}} \epsilon^{\alpha \beta \gamma} \epsilon_{ab} X\,,~~50^{\alpha \beta a}_{\sigma b} \equiv \frac{1}{\sqrt{6}} \mathbb{S}^{\alpha \beta a}_{\sigma b} + \frac{1}{12} \delta^a_b \left( \delta^\alpha_\sigma T^\beta - \delta^\beta_\sigma T^\alpha \right)\,, \nonumber \\
50^{\alpha a b}_{\sigma \rho} & \equiv & \frac{1}{2\sqrt{3}} \epsilon^{ab} {\cal S}^\alpha_{\sigma \rho}\,,~~50^{\alpha \beta a}_{bc}  \equiv  \frac{1}{2\sqrt{3}} \left(\delta^a_c \overline{\Omega}^{\alpha \beta}_b - \delta^a_b \overline{\Omega}^{\alpha \beta}_c \right)\,, \nonumber \\
50^{\alpha a b}_{\sigma c} & \equiv & \frac{1}{2\sqrt{6}} \left(\delta^b_c O^{\alpha a}_\sigma - \delta^a_c O^{\alpha b}_\sigma \right)\,,~~50^{\alpha ab}_{cd} = \frac{1}{6} \left(\delta^a_c \delta^b_d - \delta^a_d \delta^b_c \right) T^\alpha\,. \eeqa
It is straightforward to see from the above decomposition and Eq. (\ref{16_126_su5}) that the $B-L$ carrying fields ${\cal S}^\alpha_{\sigma \rho}$ and $\mathbb{S}^{\alpha \beta a}_{\sigma b}$ couple to quarks only while the $\overline{\Omega}^{\alpha \beta}_a$ has only a lepto-quark vertex. Similarly, $X$ couples to the leptons only. As a result, only $T^\alpha$ can contribute to the nucleon decays. The corresponding couplings are given by
\beqa \label{126/50}
-{\cal L}^{\overline{\bf 126}/50}_Y & \supset & -i \frac{2}{3\sqrt{5}}\, F_{AB}\, \left(u^{CT}_{\gamma A}\,C^{-1}\, e^C_B +\frac{1}{2} \epsilon_{\alpha \beta \gamma}\, \epsilon_{ab}\, q^{\alpha a T}_A\,C^{-1}\, q^{\beta b}_B \right)T_2^\gamma\,+\,{\rm h.c.}\,. \eeqa

\subsection{${\bf 16}_F$-${\bf 16}_F$-${\bf 120}_H$ couplings}
Finally, consider the Yukawa interactions with ${\bf 120}_H$ as 
\be \label{16_120}
-{\cal L}^{\bf 120}_Y = G_{AB}\,{\bf 16}^T_A\,C^{-1}\,{\bf 16}_B\,{\bf 120}_H\,+\,{\rm h.c.}\,, \ee
where $G_{AB}=-G_{BA}$ is anti-symmetric coupling unlike the previous cases. One can also have a similar but independent couplings with ${\bf 120}_H^\dagger$ if the latter is complex scalar. Under $SU(5)$, the above can be written as
\beqa \label{16_120_su5}
-{\cal L}^{\bf 120}_Y &=& i \frac{2}{\sqrt{3}}\, G_{AB}\,\Big[- 2\, 1{^T_A}\,C^{-1}\, \overline{5}_{i B}\, 5^{i}\, - 10^{ij\,T}_A\,C^{-1}\,\overline{5}{_{i B}}\,\overline{5}_j\, + \sqrt{2}\,\overline{5}{^T_{i A}}\,C^{-1}\,\overline{5}_{j B}\,10^{ij}\Big.\nonumber \\
& - & \sqrt{2}\,1{^T_{A}}\,C^{-1}\,10^{ij}_{B}\,\overline{10}_{ij}\,-\,  \frac{1}{2\sqrt{2}}\,\epsilon_{ijklm}\,10^{ij\,T}_A\,C^{-1}\,10^{mn}_B\,45^{kl}_n\,+ \sqrt{2}\,\overline{5}{^T_{i A}}\,C^{-1}\, 10^{jk}_B\,\overline{45}^i_{jk}\Big]\,+\,{\rm h.c.}\nonumber\\\eeqa

The coupling with $5$-plet involves only the SM singlet leptons. Using the decompositions of $\overline{5}$, we find that 
\beqa \label{120/5}
%-{\cal L}^{{\bf 120}/5}_Y & \supset & -i \frac{4}{\sqrt{3}}\, G_{AB}\, {\color{red} \nu^C_A \, d^C_{\alpha B}} T^\alpha\,+\,{\rm h.c.}\,, \nonumber \\
-{\cal L}^{{\bf 120}/\overline{5}}_Y & \supset & -i \frac{2}{\sqrt{3}}\, G_{AB}\, \left(\epsilon^{\alpha \beta \gamma}\, u^{CT}_{\alpha A}\,C^{-1}\, d^C_{\beta B} - \epsilon_{ab}\, q^{\gamma a T}_A\,C^{-1}\, l^b_B \right)\overline{T}_{1\gamma}\,+\,{\rm h.c.}\,. \eeqa
Like in the case of $\overline{126}_H$, the $\overline{10}$ belonging to ${\bf 120}_H$ also couples to $\nu^C$ only and hence does not play any role in the nucleon decay. The $10$-plet can be decomposed as
\beqa \label{10dec}
10^{\alpha \beta} = \overline{\Theta}^{\alpha \beta}\,,~~10^{\alpha a} = \frac{1}{\sqrt{2}} \Delta^{\alpha a}\,,~~10^{ab} = \epsilon^{ab}\,\overline{s}\,.\eeqa
Apparently, $\overline{s}$ couples to the leptons only. The field $\Delta^{\alpha a}$ has the same quantum charge as the one in Eq. (\ref{15dec}).  As it is evident from Eq. (\ref{16_120_su5}), $\overline{\Delta}$ and $\Theta$ couple to the right-handed neutrinos. The interactions with $\Delta^{\alpha a}$ and $\overline{\Theta}^{\alpha \beta}$ are obtained as
\beqa \label{120/10}
-{\cal L}^{{\bf 120}/10}_Y & \supset & -i \frac{2}{\sqrt{3}}\, G_{AB}\, \left(2\epsilon_{ab}\, d^{CT}_{\alpha A}\,C^{-1}\, l^a_B\, \Delta^{\alpha b}\,- \sqrt{2}\,d^{CT}_{\alpha A}\,C^{-1}\,d^{C}_{\beta B}\,\overline{\Theta}^{\alpha \beta}\right)\,+\,{\rm h.c.}\,. \eeqa
The first term is similar to Eq. (\ref{126/15}) which has only lepto-quark couplings. $\overline{\Theta}^{\alpha \beta}$ has only the diquark couplings. Therefore, none of these scalars residing in $10$ and $\overline{10}$ can induce proton decay through $d=6$ operators. However, they can contribute to $B-L$ violating decays of nucleon as we discuss later.

The couplings of remaining fields residing in the $45$- and $\overline{45}$-plets can straightforwardly be computed using the decomposition already given in Eq. (\ref{45dec}). We find

\beqa \label{120/45}
-{\cal L}^{{\bf 120}/45}_Y & \supset & i \frac{2}{\sqrt{3}}\, G_{AB}\, \Big(2\, u^{CT}_{\alpha A}\,C^{-1}\,e^C_B\,T^\alpha\,+\,\epsilon^{\alpha \beta \gamma}\,u^{CT}_{\alpha A}\,C^{-1}\,u^C_{\beta B}\,\overline{{\cal T}}_\gamma \Big. \nonumber\\ 
& + & \Big. \epsilon_{\alpha \beta \gamma}\,  \epsilon_{ab}\, q^{\alpha a T}_A\,C^{-1}\,q^{\beta c}_B\, \mathbb{T}^{b \gamma}_c\,-\,\epsilon_{\alpha \beta \gamma}\,e^{C  T}_A\,C^{-1}\,q^{\gamma a}_B\,\overline{\Omega}^{\alpha \beta}_a \Big)\,+{\rm h.c.}\,. \eeqa
\beqa \label{120/45b}
-{\cal L}^{{\bf 120}/\overline{45}}_Y & \supset & -i \frac{2}{\sqrt{3}}\, G_{AB}\, \left[\left(\epsilon^{\alpha \beta \gamma}\, u^{CT}_{\alpha A}\,C^{-1}\,d^C_{\beta B} + \epsilon_{ab}\, q^{\gamma a T}_A\,C^{-1}\,l^b_B\right) \overline{T}_{2 \gamma}\, \right. \nonumber \\ 
& + &  2\, e^{CT}_A\,C^{-1}\, d^C_{\alpha B}\, {\cal T}^\alpha\,+\, 2\, \epsilon_{bc}\,q^{\alpha a T}_A\,C^{-1}\,l^b_B\,\overline{\mathbb{T}}^c_{a \alpha} \nonumber \\
& - & \left. \epsilon^{\alpha \beta \gamma}\,\epsilon_{ab}\,l^{bT}_A\,C^{-1}\,u^C_{\gamma B}\,\Omega^a_{\alpha\beta} \right]\,+{\rm h.c.}\,. \eeqa
Note that the triplet residing in $45$ of $SU(5)$ does not couple to the quarks because of flavour anti-symmetric couplings. Moreover, unlike in Eq. (\ref{126/45}), the fields ${\cal T}^\alpha$ and $\mathbb{T}^{a \alpha}_b$ have quark-quark couplings. Their conjugate fields have lepto-quark couplings. $\Omega^a_{\alpha\beta}$ and its conjugate partner have only the lepto-quark couplings and therefore they do not induce $d=6$ operator but contribute to the $d=7$ operators.

All together the couplings in Eqs. (\ref{10/5},\ref{126/5},\ref{126/45},\ref{126/50},\ref{120/5},\ref{120/45},\ref{120/45b}) determine the amplitude of nucleon decays in the most general renormalizable versions of the $SO(10)$ GUTs. The noteworthy features are:
\begin{itemize}
\item Out of the six pairs of colour triplets with different electroweak charges (see Table \ref{tab:scalars}), only three pairs have couplings suitable for destabilizing protons at the tree level through $d=6$ operators. They are $T$-$\overline{T}$, ${\cal T}$-$\overline{\cal T}$ and $\mathbb{T}$-$\overline{\mathbb{T}}$. The remaining three pairs, namely $\Theta$-$\overline{\Theta}$, ${\Delta}$-$\overline{\Delta}$ and $\Omega$-$\overline{\Omega}$ can induce nucleon decay through $d=7$ operators.
\item If the underlying model does not contain ${\bf 120}_H$ then nucleon decay is induced only by pair of colour triplets $T$, $\overline{T}$ with $|Y|=1/3$. The ${\bf 10}_H$ contains a pair of such fields while $\overline{\bf 126}_H$ contains a pair and additional $T$. While the other two triplets are present in $\overline{\bf 126}_H$, they have only the lepto-quark couplings. Moreover, $\Delta$ together with $\overline{T}$ can also induce nucleon decay through $B-L$ breaking.
\item In the presence of ${\bf 120}_H$, the nucleon decays can also receive contributions from ${\cal T}$-$\overline{\cal T}$ and $\mathbb{T}$-$\overline{\mathbb{T}}$. These fields can contribute by themselves as well as by mixing with the fields of same charges residing in $\overline{\bf 126}_H$ in the models in which both $\overline{\bf 126}_H$ and ${\bf 120}_H$ are present.
\end{itemize}
The results obtained in this section can also be used to derive the couplings relevant for proton decays in the most general renormalizable $SU(5)$ GUT models.

\section{Dimension-6 effective operators}
\label{sec:operators_d6}
We now compute the leading $d=6$ effective operators relevant for the $B$ and $L$ violating decays of baryons by integrating out the relevant scalar fields. We treat each case of ${\bf 10}_H$, $\overline{\bf 126}_H$ and ${\bf 120}_H$ separately and comment on the possibility of mixing between the scalar fields at the end of this section.

\subsection{From ${\bf 10}$-plet}
The contribution to the nucleon decay in models with only ${\bf 10}_H$ arises from a pair of color triplet and anti-triplet, $T$ and $\overline{T}$ , as can be seen from Eq. (\ref{10/5}). The most general mass terms for these triplets can be written as:
\be \label{triplet_mix}
m_T^2\, T^\dagger T + m_{\overline{T}}^2\, \overline{T}^\dagger \overline{T} + \left( \mu^2\, T\, \overline{T}+{\rm h.c.}\right)\,\ee
where the first two terms can originate from $SO(10)$ invariant combination ${\bf 10}_H^\dagger {\bf 10}_H$ and the last term from ${\bf 10}_H^2$ \cite{Slansky:1981yr}. In general, $m_T \neq m_{\overline{T}}$ as the mass splitting can arise if the model also contains ${\bf 45}_H$ and/or ${\bf 54}_H$. Because of the presence of the mixing term, the physical states are different and can be obtained by the following replacements:
\be \label{T_redf}
T \to T \cos\theta_T + \overline{T}^* \sin\theta_T\,,~~ \overline{T} \to \overline{T} \cos\theta_T - T^* \sin\theta_T\,, \ee
Here, the mixing angle is obtained from Eq. (\ref{triplet_mix}) as 
\be \label{mixing_angle}
\tan 2\theta_T = \frac{2 \mu^2}{m_{\overline{T}}^2 - m_T^2}\,. \ee 
With the above replacements, Eq. (\ref{triplet_mix}) becomes
\be \label{mass_term_T}
M_T^2\, T^\dagger T + M_{\overline{T}}^2\, \overline{T}^\dagger \overline{T}\,.\ee

Substituting Eq. (\ref{T_redf}) in Eq. (\ref{10/5}) and integrating out the physical triplet and anti-triplet, we obtain the following effective Lagrangian relevant for $B$ violating baryon decays after some straight-forward algebraic manipulations:
\beqa \label{op_10_0}
{\cal L}^{{\bf 10}}_{\rm eff} &=& 8 \left(\frac{c_T^2}{M_T^2}+\frac{s_T^2}{M_{\overline{T}}^2}\right) H_{AB} H^\dagger_{CD}\,\, \epsilon^{\alpha \beta \gamma}\, u_{\alpha A}^T C^{-1} d_{\beta B}\, {e^C_C}^\dagger C^{-1}{u^C_{\gamma D}}^* \nonumber \\
& + &8 \left(\frac{s_T^2}{M_T^2}+\frac{c_T^2}{M_{\overline{T}}^2}\right) H_{AB} H^\dagger_{CD}\,\, \epsilon^{\alpha \beta \gamma}\, u{^{CT}_{\alpha A}} C^{-1} d_{\beta B}^C\, \left(e_C^\dagger C^{-1} u_{\gamma D}^* - \nu_C^\dagger C^{-1} d_{\gamma D}^*\right)  \nonumber \\
& + &  8 s_T c_T \left(\frac{1}{M_T^2}-\frac{1}{M_{\overline{T}}^2} \right) H_{AB} H^T_{CD}\,\,\epsilon^{\alpha \beta \gamma}\, u{^{CT}_{\alpha A}} C^{-1} d^C_{\beta B}\, {e^{CT}_C} C^{-1} {u^C_{\gamma D}}\nonumber \\
& + & 8 s_T c_T \left(\frac{1}{M_T^2}-\frac{1}{M_{\overline{T}}^2} \right) H_{AB} H^T_{CD}\,\, \epsilon^{\alpha \beta \gamma}\, u_{\alpha A}^T C^{-1} d_{\beta B}\, \left(e_C^T C^{-1} u_{\gamma D} - \nu_C^T C^{-1} d_{\gamma D} \right)\nonumber \\
&+& {\rm h.c.}\,,\eeqa
where $c_T=\cos\theta_T$ and $s_T=\sin\theta_T$. We have used the identity $C^\dagger = C^T= C^{-1} =-C$ in determining Eq. (\ref{op_10_0}).

The effective operators in the physical basis of fermions can be obtain by replacing $f \to U_f f$ in the above ${\cal L}^{{\bf 10}}_{\rm eff}$. The unitary matrices $U_f$ and $U_{f^C}$ (with $f=u,d,e,\nu$) can be explicitly computed from the corresponding fermion mass matrices. In the physical basis, we get
\beqa \label{op_10}
{\cal L}^{{\bf 10}}_{\rm eff} &=& h[u_A,d_B,e^C_C,u^C_D]\,\, \epsilon^{\alpha \beta \gamma}\, u_{\alpha A}^T\,C^{-1}\,d_{\beta B}\, {e^C_C}^\dagger \,C^{-1}\,u{^{C*}_{\gamma D}} \nonumber \\
& + & h[u^C_A,d^C_B,e_C,u_D]\,\, \epsilon^{\alpha \beta \gamma}\, u{{^{CT}_{\alpha A}}}\,C^{-1}\, d_{\beta B}^C\, e_C^\dagger \,C^{-1}\, u_{\gamma D}^*\, \nonumber \\
& + & h[u^C_A,d^C_B,\nu_C,d_D]\,\, \epsilon^{\alpha \beta \gamma}\, u{^{CT}_{\alpha A}}\, C^{-1}\, d_{\beta B}^C\, \nu_C^\dagger\,C^{-1}\,d_{\gamma D}^*\, \nonumber\\
& + & h^\prime[u^C_A,d^C_B,e^C_C,u^C_D]\,\, \epsilon^{\alpha \beta \gamma}\, u{^{CT}_{\alpha A}}\,C^{-1}\, d^C_{\beta B}\, e{^{CT}_C}\,C^{-1}\,u^C_{\gamma D} \nonumber \\
& + & h^\prime[u_A,d_B,e_C,u_D]\,\, \epsilon^{\alpha \beta \gamma}\, u_{\alpha A}^T\,C^{-1}\,d_{\beta B}\, e_C^T\,C^{-1}\,u_{\gamma D}\, \nonumber \\
& + & h^\prime[u_A,d_B,\nu_C,d_D]\,\, \epsilon^{\alpha \beta \gamma}\, u_{\alpha A}^T\,C^{-1}\,d_{\beta B}\, \nu_C^T\,C^{-1}\,d_{\gamma D}\,\, +{\rm h.c.}\,,\eeqa
with
\beqa \label{coff_op_10}
h[u_A,d_B,e^C_C,u^C_D] &=& 8 \left(\frac{c_T^2}{M_T^2}+\frac{s_T^2}{M_{\overline{T}}^2}\right)  \left(U_u^T H U_d \right)_{AB}\,\left(U_{e^C}^\dagger H^\dagger U_{u^C}^*\right)_{CD}\,, \nonumber \\
h[u^C_A,d^C_B,e_C,u_D] &=& 8 \left(\frac{s_T^2}{M_T^2}+\frac{c_T^2}{M_{\overline{T}}^2}\right)  \left(U_{u^C}^T H U_{d^C}\right){_{AB}}\, \left(U_{e}^\dagger H^\dagger U_{u}^* \right){_{CD}}\,, \nonumber \\
h[u^C_A,d^C_B,\nu_C,d_D] &=& - 8 \left(\frac{s_T^2}{M_T^2}+\frac{c_T^2}{M_{\overline{T}}^2}\right) \left(U_{u^C}^T H U_{d^C} \right){_{AB}}\,\left(U_{\nu}^\dagger H^\dagger U_{d}^*\right){_{CD}}\,, \nonumber \\
h^\prime[u^C_A,d^C_B,e^C_C,u^C_D] &=& 8 s_T c_T \left(\frac{1}{M_T^2}-\frac{1}{M_{\overline{T}}^2} \right)  \left(U_{u^C}^T H U_{d^C} \right){_{AB}}\,\left(U_{e^C}^T H^T U_{e^D}\right){_{CD}}\,, \nonumber \\
h^\prime[u_A,d_B,e_C,u_D] &=& 8 s_T c_T \left(\frac{1}{M_T^2}-\frac{1}{M_{\overline{T}}^2} \right) \left(U_u^T H U_d \right)_{AB}\,\left(U_e^T H^T U_u\right)_{CD}\,, \nonumber \\
h^\prime[u_A,d_B,\nu_C,d_D] &=&-8 s_T c_T \left(\frac{1}{M_T^2}-\frac{1}{M_{\overline{T}}^2} \right) \left(U_u^T H U_d \right)_{AB}\,\left(U_\nu^T H^T U_d \right)_{CD}\,. \eeqa
In the absence of mixing term between $T$ and $\overline{T}$ (i.e. $\theta_T = 0$), all $h^\prime = 0$. Moreover, the coefficients of the first and the next two operators become independent from each other.

\subsection{From $\overline{\bf 126}$-plet}
Unlike in the case of ${\bf 10}_H$, the color triplets and anti-triplet do not mix with each other in this since $(\overline{\bf 126}_H)^2$ is forbidden by $SO(10)$ gauge symmetry \cite{Slansky:1981yr}. The triplets residing in $\overline{5}$ and $50$ belonging to $\overline{\bf 126}_H$ can mix with each other in general but such a mixing can arise from the gauge invariant interactions with other fields and hence is model dependent.  Further, as noted earlier, ${\cal T}^\alpha$ and $\overline{\mathbb{T}}^c_{a \alpha}$ have only the lepto-quark couplings. Therefore, the effective operators relevant for the nucleon decay take simple form in case of $\overline{\bf 126}_H$. Integrating out the triplets and ant-triplet from Eqs.(\ref{126/5},\ref{126/45},\ref{126/50}), we find 
\beqa \label{op_126_0}
{\cal L}^{\overline{\bf 126}}_{\rm eff} &=& \frac{2}{45}\left(\frac{1}{M_{T_1}^2}-\frac{2}{M_{T_2}^2}\right)\,F_{AB} F^\dagger_{CD}\,\, \epsilon^{\alpha \beta \gamma}\, u_{\alpha A}^T C^{-1} d_{\beta B}\, {e^C_C}^\dagger C^{-1}{u^C_{\gamma D}}^* \nonumber \\
& - &\frac{2}{15}\frac{1}{M_{\overline{T}}^2}\,F_{AB} F^\dagger_{CD}\,\, \epsilon^{\alpha \beta \gamma}\, u{^{CT}_{\alpha A}} C^{-1} d_{\beta B}^C\, \left(e_C^\dagger C^{-1} u_{\gamma D}^* - \nu_C^\dagger C^{-1} d_{\gamma D}^*\right)\, +{\rm h.c.}\,,\eeqa
where $M_{T_i}$ and $M_{\overline{T}}$ are masses of triplets $T_i$ and anti-triplet $\overline{T}$, respectively.

In the physical basis, the above can be rewritten as
\beqa \label{op_126}
{\cal L}^{\overline{\bf 126}}_{\rm eff} &=& f[u_A,d_B,e^C_C,u^C_D]\,\, \epsilon^{\alpha \beta \gamma}\, u_{\alpha A}^T\,C^{-1}\,d_{\beta B}\, {e^C_C}^\dagger\,C^{-1}\,u{^{C*}_{\gamma D}} \nonumber \\
& + & f[u^C_A,d^C_B,e_C,u_D]\,\, \epsilon^{\alpha \beta \gamma}\, u{^{CT}_{\alpha A}}\,C^{-1}\,d_{\beta B}^C\, e_C^\dagger\,C^{-1}\,u_{\gamma D}^*\, \nonumber \\
& + & f[u^C_A,d^C_B,\nu_C,d_D]\,\, \epsilon^{\alpha \beta \gamma}\, u{^{CT}_{\alpha A}}\,C^{-1}\,d_{\beta B}^C\, \nu_C^\dagger\,C^{-1}\,d_{\gamma D}^*\, +{\rm h.c.}\,,\eeqa
with
\beqa \label{coff_op_126}
f[u_A,d_B,e^C_C,u^C_D] &=& \frac{2}{45}\left(\frac{1}{M_{T_1}^2}-\frac{2}{M_{T_2}^2}\right) \left(U_u^T F U_d \right)_{AB}\,\left(U_{e^C}^\dagger F^\dagger U_{u^C}^*\right)_{CD}\,, \nonumber \\
f[u^C_A,d^C_B,e_C,u_D] &=& -\frac{2}{15}\frac{1}{M_{\overline{T}}^2} \left(U_{u^C}^T F U_{d^C}\right){_{AB}}\, \left(U_{e}^\dagger F^\dagger U_{u}^* \right){_{CD}}\,, \nonumber \\
f[u^C_A,d^C_B,\nu_C,d_D]&=&  \frac{2}{15}\frac{1}{M_{\overline{T}}^2} \left(U_{u^C}^T F U_{d^C} \right){_{AB}}\,\left(U_{\nu}^\dagger F^\dagger U_{d}^*\right){_{CD}}\,. \eeqa
The structure of these operators is same as of the ones obtained in case of ${\bf 10}_H$ with a noteworthy difference of relative factor between the contributions mediated by color triplets and anti-triplet.

\subsection{From ${\bf 120}$-plet}
Since $({\bf 120}_H)^2$ is a gauge invariant, $T \in 45$ and $\overline{T}_2 \in \overline{45}$ can have a mixing term. Similarly, ${\cal T}$ and $\mathbb{T}^{a \alpha}_b$ can mix with their conjugate partners appearing in Eqs. (\ref{120/45},\ref{120/45b}). Adopting the similar procedure discussed in the first subsection, we obtain the following operators after eliminating the different colour triplet scalars:
\beqa \label{op_120_0}
{\cal L}^{{\bf 120}}_{\rm eff} &=& \frac{4}{3}\left(\frac{1}{M_{\overline{T}_1}^2}-\frac{c_T^2}{M_{\overline{T}_2}^2} - \frac{s_T^2}{M_T^2}\right)\,G_{AB} G^\dagger_{CD}\,\, \epsilon^{\alpha \beta \gamma}\, u{^{CT}_{\alpha A}} \,C^{-1}\,d_{\beta B}^C\, \left(e_C^\dagger\,C^{-1}\,u_{\gamma D}^* - \nu_C^\dagger C^{-1} d_{\gamma D}^*\right)\, \nonumber \\
& + & \frac{8}{3}s_T c_T \left(\frac{1}{M{^2_T}}-\frac{1}{M_{\overline{T}_2}^2} \right) G_{AB} G_{CD}\,\,\epsilon^{\alpha \beta \gamma}\, u{^{CT}_{\alpha A}}\,C^{-1}\,d^C_{\beta B}\, e{^{CT}_C}\,C^{-1}\,{u^C_{\gamma D}} \nonumber \\
& -& \frac{8}{3} s_{\cal T} c_{\cal T} \left(\frac{1}{M_{\cal T}^2}-\frac{1}{M_{\overline{\cal T}}^2} \right) G_{AB} G_{CD}\,\,\epsilon^{\alpha \beta \gamma}\, u{^{CT}_{\alpha A}}\,C^{-1}\,u^C_{\beta B}\, e{^{CT}_C}\,C^{-1}\,{d^C_{\gamma D}}\, \nonumber \\
& -& \frac{8}{3} s_{\mathbb{T}} c_{\mathbb{T}} \left(\frac{1}{M_\mathbb{T}^2}-\frac{1}{M_{\overline{\mathbb{T}}}^2} \right) G_{AB} G_{CD}\,\epsilon^{\alpha \beta \gamma}\, \epsilon_{bc} \epsilon_{da}\,q_{\alpha A}^{a T}\,C^{-1}\,l^b_B\, q_{\beta C}^{d T}\,C^{-1}\, q^c_{\gamma D}\,+\, {\rm h.c.}\,\eeqa
Here, $\theta_T$, $\theta_{\cal T}$ and $\theta_{\mathbb{T}}$ are angles which parametrize mixing term between $T$-$\overline{T}$, ${\cal T}$-$\overline{\cal T}$ and $\mathbb{T}$-$\overline{\mathbb{T}}$, respectively, in an analogous way to Eq. (\ref{mixing_angle}). Since $T$ has only the lepto-quark couplings, its contribution to the nucleon decays arises only through the mixing term.

Using the Fierz rearrangement
\be \label{fierz}
\left(\psi_1^T C^{-1} \psi_2 \right) \left(\psi_3^T C^{-1} \psi_4 \right) = - \left(\psi_1^T C^{-1} \psi_3 \right) \left(\psi_4^T C^{-1} \psi_2 \right) -\left(\psi_1^T C^{-1} \psi_4 \right) \left(\psi_2^T C^{-1} \psi_3 \right)\,,\ee
we rewrite the operator given in the third line in Eq. (\ref{op_120_0}) as
\be \label{3op}
 G_{AB} G_{CD}\,\epsilon^{\alpha \beta \gamma}\, u{^{CT}_{\alpha A}}\,C^{-1}\, u^C_{\beta B}\, e{^{CT}_C}\,C^{-1}\,d^C_{\gamma D} = 2 G_{AD} G_{CB}\,\,\epsilon^{\alpha \beta \gamma}\, u{^{CT}_{\alpha A}}\,C^{-1}\,d^C_{\beta B}\, e{^{CT}_C}\,C^{-1}\,u^C_{\gamma D}\,,\ee
where we also use $G_{AB} = -G_{BA}$. Further, using $\epsilon_{bc} \epsilon_{da} = \delta_{bd} \delta_{ca} - \delta_{ba} \delta_{cd}$ and Eq. (\ref{fierz}) the fourth operator can be simplified to
\beqa \label{4op}
& & G_{AB} G_{CD} \epsilon^{\alpha \beta \gamma} \epsilon_{bc} \epsilon_{da}\, q_{\alpha A}^{a T}\,C^{-1}\,l^b_B\, q_{\beta C}^{d T}\,C^{-1}\,q^c_{\gamma D}\nonumber \\
&=& - \left(G_{AB} G_{CD} - 2 G_{AD} G_{CB}\right)\, \epsilon^{\alpha \beta \gamma}\, u_{\alpha A}^T\,C^{-1}\,d_{\beta B}\, e_C^T\,C^{-1}\,u_{\gamma D} \nonumber \\
&-& \left(G_{AB} G_{CD} - 2 G_{AC} G_{BD}\right)\, \epsilon^{\alpha \beta \gamma}\, u_{\alpha A}^T\,C^{-1}\,d_{\beta B}\, \nu_C^T\,C^{-1}\,d_{\gamma D}\,. \eeqa

Substituting Eqs. (\ref{3op},\ref{4op}) in Eq. (\ref{op_120_0}) and using the physical basis for fermions, we get
\beqa \label{op_120}
{\cal L}^{{\bf 120}}_{\rm eff} &=& g[u^C_A,d^C_B,e_C,u_D]\,\, \epsilon^{\alpha \beta \gamma}\, u{^{CT}_{\alpha A}}\,C^{-1}\,d_{\beta B}^C\, e_C^\dagger\,C^{-1}\, u_{\gamma D}^*\, \nonumber \\
& + & g[u^C_A,d^C_B,\nu_C,d_D]\,\, \epsilon^{\alpha \beta \gamma}\, u{^{CT}_{\alpha A}}\,C^{-1}\,d_{\beta B}^C\, \nu_C^\dagger\,C^{-1}\,d_{\gamma D}^*\, \nonumber\\
& + & g^\prime[u^C_A,d^C_B,e^C_C,u^C_D]\,\, \epsilon^{\alpha \beta \gamma}\, u{^{CT}_{\alpha A}}\,C^{-1}\,d_{\beta B}^C\, e{^{CT}_C}\,C^{-1}\,u^C_{\gamma D}\,\nonumber \\
& + & g^\prime[u_A,d_B,e_C,u_D]\,\, \epsilon^{\alpha \beta \gamma}\, u_{\alpha A}^T\,C^{-1}\,d_{\beta B}\, e_C^T\,C^{-1}\,u_{\gamma D}\, \nonumber \\
& + & g^\prime[u_A,d_B,\nu_C,d_D]\,\, \epsilon^{\alpha \beta \gamma}\, u_{\alpha A}^T\,C^{-1}\,d_{\beta B}\, \nu_C^T\,C^{-1}\,d_{\gamma D}\,\, +{\rm h.c.}\,,\eeqa
with
\beqa \label{coff_op_120}
g[u^C_A,d^C_B,e_C,u_D] &=& \frac{4}{3}\left(\frac{1}{M_{\overline{T}_1}^2}-\frac{c_T^2}{M_{\overline{T}_2}^2}- \frac{s_T^2}{M_T^2}\right)\, \left(U_{u^C}^T G U_{d^C}\right){_{AB}}\, \left(U_{e}^\dagger G^\dagger U_{u}^* \right){_{CD}}\,, \nonumber \\
g[u^C_A,d^C_B,\nu_C,d_D] &=& -\frac{4}{3}\left(\frac{1}{M_{\overline{T}_1}^2}-\frac{c_T^2}{M_{\overline{T}_2}^2}- \frac{s_T^2}{M_T^2}\right)\, \left(U_{u^C}^T G U_{d^C} \right){_{AB}}\,\left(U_{\nu}^\dagger G^\dagger U_{d}^*\right){_{CD}}\,,\nonumber \\
g^\prime[u^C_A,d^C_B,e^C_C,u^C_D] &=& \frac{8}{3} s_T c_T \left(\frac{1}{M_T^2}-\frac{1}{M_{\overline{T}_2}^2} \right)\, \left(U_{u^C}^T G U_{d^C} \right){_{AB}}\,\left(U_{e^C}^T G U_{u^C}\right){_{CD}}\, \nonumber \\
& - & \frac{16}{3} s_{\cal T} c_{\cal T} \left(\frac{1}{M_{\cal T}^2}-\frac{1}{M_{\overline{\cal T}}^2} \right) \left(U_{u^C}^T G U_{u^C} \right)_{AD}\,\left(U_{e^c}^T G U_{d^C}\right)_{CB}\,, \nonumber \\
g^\prime[u_A,d_B,e_C,u_D] &=& \frac{8}{3} s_{\mathbb{T}} c_{\mathbb{T}} \left(\frac{1}{M_\mathbb{T}^2}-\frac{1}{M_{\overline{\mathbb{T}}}^2} \right) \Big[\left(U_u^T G U_d \right)_{AB}\,\left(U_e^T G U_u\right)_{CD} \big. \nonumber \\ 
&-& \Big. 2 \left(U_u^T G U_u \right)_{AD}\,\left(U_e^T G U_d\right)_{CB}\Big]\,, \nonumber \\
g^\prime[u_A,d_B,\nu_C,d_D] &=& \frac{8}{3} s_{\mathbb{T}} c_{\mathbb{T}} \left(\frac{1}{M_\mathbb{T}^2}-\frac{1}{M_{\overline{\mathbb{T}}}^2} \right) \Big[\left(U_u^T G U_d \right)_{AB}\,\left(U_\nu^T G U_d\right)_{CD} \big. \nonumber \\ 
&-& \Big. 2 \left(U_u^T G U_\nu \right)_{AC}\,\left(U_d^T G U_d\right)_{BD}\Big]\,. \eeqa
The structure of effective couplings $g^\prime$ are very different from $h^\prime$ due to additional contributions supplied by the new triplets.

Altogether, the operators listed in Eqs. (\ref{op_10},\ref{op_126},\ref{op_120}) quantify the $B$ and $L$ violating but $B-L$ conserving baryon decays mediated by scalars residing in ${\bf 10}_H$, $\overline{\bf 126}_H$ and ${\bf 120}_H$, respectively, where we have also considered the possibility of the mixing between the various scalars arising from the given representation. In general, the scalar fields of the same SM charges belonging to different GUT representations can also mix. For example, in the models with at least two of these scalar representations, different triplets and anti-triplets can mix and the lightest pair would be a particular linear combination of them. Since the contribution of the lightest pair is expected to be the most dominant, the operators obtained by integrating out this pair would be the most relevant in quantifying the proton decays. In general, such operators can be obtained with coefficients which are linear combinations of the relevant $h$, $f$ and $g$, for example. However, the exact quantification of these mixing depends not only on the specification of the Yukawa sector but also on the full scalar potential of the underlying model. Therefore, this exercise is highly model-dependent. Nevertheless, the results obtained above can straightforwardly be used to compute proton decay in specific models in which mixing between the various triplets and anti-triplets is deterministic.

\section{Dimension-7 effective operators}
\label{sec:operators_d7}
In the previous section, we consider the leading order $d=6$ operators which violate $B$ and $L$ but conserve $B-L$. A new class of operators which violate $B-L$ by two units arise at $d=7$ \cite{Weinberg:1980bf,Weldon:1980gi}. Such operators are induced by quartic couplings involving the SM singlet field $\sigma$, one of the electroweak doublets and two color triplet fields. Using the fields listed in Table \ref{tab:scalars}, the  following invariants can be constructed for this kind of quartic term:
\beqa 
\sigma\, D^a\,T^\alpha\,\overline{\Delta}_{\alpha a}\, &,&~~\sigma\, \overline{D}_a\,\overline{T}_\alpha\,\Delta^{\alpha a}\,; \label{q1} \\
\sigma\, D^a\,\Theta_{\alpha \beta}\,\overline{\Omega}^{\alpha \beta}_a\, &,&~~\sigma\, \overline{D}_a\,\overline{\Theta}^{\alpha \beta}\,\Omega_{\alpha \beta}^a\,; \label{q2} \\
\sigma\, D^a\,\overline{\Theta}^{\alpha \beta}\,\Delta^{\gamma b}\,\epsilon_{\alpha \beta \gamma}\,\epsilon_{ab}\, &,&~~\sigma\, \overline{D}_a\,\Theta_{\alpha \beta}\,\overline{\Delta}_{\gamma b}\,\epsilon^{\alpha \beta \gamma}\,\epsilon^{ab}\,; \label{q3} \\
\sigma\, D^a\,\overline{\Delta}_{\alpha b}\,\mathbb{T}^{\alpha b}_a\, &,&~~\sigma\, \overline{D}_a\,\Delta^{\alpha b}\,\overline{\mathbb{T}}_{\alpha b}^a\,. \label{q4} \eeqa
When $\sigma$ acquires a VEV, $B-L$ symmetry is broken and the above quartic terms can give rise to $d=7$ operators involving four fermions and a Higgs doublet. After the electroweak symmetry breaking, this generates effective four-fermion operators with $\Delta (B-L) = -2$. Because of the latter, they give rise to novel decay channels for the proton and neutron \cite{Babu:2012iv,Babu:2012vb}.

The quartic terms listed in Eqs. (\ref{q1}-\ref{q4}) can arise from various $SO(10)$ invariant combinations of the scalar irreps, for example $(\overline{\bf 126}_H)^4$, $(\overline{\bf 126}_H)^2 ({\bf 120}_H)^2$, etc. Such a term must contain atleast one $\overline{\bf 126}_H$ (or ${\bf 126}_H$) as a source of $\sigma$ while the electroweak doublets can come from either of ${\bf 10}_H$, $\overline{\bf 126}_H$ or ${\bf 120}_H$. Various triplets appearing in Eqs. (\ref{q1}-\ref{q4}) can come from ${\bf 10}_H$, $\overline{\bf 126}_H$, ${\bf 120}_H$ or even from the scalars like ${\bf 126}_H$ which are not part of the Yukawa sector. Although the latter does not directly couple to quarks and leptons but can still induce nucleon decay by mixing with the scalar sub-multiplets residing in ${\bf 10}_H$, $\overline{\bf 126}_H$ and ${\bf 120}_H$. The determination of exact operators in this case, therefore, requires complete specification of the scalar sector of the underlying model beyond the ones which take part into the Yukawa interactions. As we remain agnostic about the complete model in this study, we perform the subsequent analysis assuming that the color triplets fields in Eqs. (\ref{q1}-\ref{q4}) arise from either of ${\bf 10}_H$, $\overline{\bf 126}_H$ or ${\bf 120}_H$.

\subsection{From ${\bf 10}$-plet}
The generation of $d=7$ operators requires presence of atleast two different kinds of color triplets as it can be seen from  Eqs. (\ref{q1}-\ref{q4}). Since ${\bf 10}_H$ contains only one kind of color triplets, it cannot give rise to such operators at the leading order by itself.

\subsection{From $\overline{\bf 126}$-plet}
As it can be seen from Table \ref{tab:scalars}, $\overline{\bf 126}_H$ does not contain $\overline{\Theta}^{\alpha \beta}$ and $\mathbb{T}^{a \alpha}_b$. Among the remaining fields responsible for generating $d=7$ operators, $\Theta_{\alpha \beta}$ and $\overline{\Delta}_{\alpha a}$ reside in $\overline{10}$ of $SU(5)$ which couple to the RH neutrinos only, see Eq. (\ref{16_126_su5}). Moreover, $\Delta^{\alpha a}$, $\overline{\Omega}^{\alpha \beta}_a$ and $\overline{\mathbb{T}}^a_{\alpha b}$ have only the lepto-quark couplings. Therefore, only the second term in Eq. (\ref{q1}) can induce the required operator. Using Eqs. (\ref{126/15},\ref{126/45}) and integrating out $\Delta^{\alpha a}$ and $\overline{T}_\alpha$, we get 
\beqa \label{op_126_0_d7}
{\cal L}^{\overline{\bf 126}}_{\rm eff} &=& \frac{4 \lambda v_\sigma}{15 M_\Delta^2 M_{\overline{T}}^2}\,F_{AB}^*F_{CD}^*\,\epsilon^{\alpha \beta \gamma}\, \epsilon^{ab}\overline{D}_a\,u^{C \dagger}_{\alpha A}\,C^{-1}\,{d^C_{\beta B}}^*\, l_{b C}^\dagger\,C^{-1}\,{d^C_{\gamma D}}^*\,+\,{\rm h.c.}\,,\eeqa
where $\lambda$ is a coupling of the quartic term and $v_\sigma$ is the VEV that breaks $B-L$. The above can be identified with operator $\tilde{O}_1$ listed in \cite{Babu:2012vb}\footnote{Note that we have assumed absence of ${\bf 126}_H$ which forbids $T$-$\overline{T}$ mixing as it was also the case while deriving $d=6$ operators earlier. If such mixing is allowed, one also finds an operator similar to $\tilde{O}_3$ given in \cite{Babu:2012vb}}.

After the electroweak symmetry breaking, the above operator reduces to $d=6$ operator containing three quarks and a neutrino field. In the physical basis, it can be parametrized as
\beqa \label{op_126_d7}
{\cal L}^{\overline{\bf 126}}_{\rm eff} &=& \tilde{f}[u^C_A,d^C_B,\nu_C,d^C_D]\,\epsilon^{\alpha \beta \gamma}\,u^{C T}_{\alpha A}\,C^{-1}\,{d^C_{\beta B}}\, \nu_C^T\,C^{-1}\,{d^C_{\gamma D}}\,+\,{\rm h.c.}\,,\eeqa
with 
\beqa \label{coff_op_126_d7}
 \tilde{f}[u^C_A,d^C_B,\nu_C,d^C_D] &=& -\frac{4 \lambda v_\sigma v_{\overline{D}}}{15 M_\Delta^2 M_{\overline{T}}^2}\, \left(U_{u^C}^T F U_{d^C} \right)_{AB}\,\left(U_\nu^T F U_{d^C}\right)_{CD}\,, \eeqa
 where $v_{\overline{D}}$ is a VEV of $\overline{D}$ residing in $\overline{\bf 126}$.

\subsection{From ${\bf 120}$-plet}
In comparison to the previous two, ${\bf 120}_H$ offers more variety of $d=7$ operators because of the mixing term allowed between various scalar sub-multiplets and their conjugates.  All four invariants listed in Eqs. (\ref{q1}-\ref{q4}) give rise to $B-L$ violating operators. Using Eqs. (\ref{q1}-\ref{q4}) and couplings for various triplet fields evaluated in section \ref{sec:couplings}, we find the following leading order operators after some straight-forward computation:
\beqa \label{op_120_0_d7}
{\cal L}^{\bf 120}_{\rm eff} &=& -\frac{8 \lambda v_\sigma}{3}\left[\left( \frac{1}{M{^2_{\Delta}}\,M{^2_{\overline{T}_1}}}+\frac{c_\Delta c_T}{M_\Delta^2 M^2_{\overline{T}_2}}\right) + \frac{s_\Delta s_T}{M_{\overline{\Delta}}^2 M_{T}^2}\right]*\nonumber\\
&& \hspace{10 pt}\, G_{AB}^*G_{CD}^*\,\epsilon^{\alpha \beta \gamma}\,\epsilon^{ab}\,\overline{D}_a \,u^{C \dagger}_{\alpha A}\,C^{-1}\,{d^C_{\beta B}}^*\,l^\dagger_{bC}\,C^{-1}\, {d^{C}_{\gamma D}}^*\, \nonumber \\
& - & \frac{4\sqrt{2} \lambda v_\sigma}{3}\left(\frac{c_\Theta c_\Omega}{M_{\overline{\Theta}}^2 M_\Omega^2} + \frac{s_\Theta s_\Omega}{M_\Theta^2 M_{\overline{\Omega}}^2} \right)\, G_{AB}^*G_{CD}^*\,\epsilon^{\alpha \beta \gamma}\, \epsilon^{ab}\,\overline{D}_a\,d^{C \dagger}_{\alpha A}\,C^{-1}\,{d^C_{\beta B}}^*\, l^\dagger_{b C}\,C^{-1}\,{u^C_{\gamma D}}^*\, \nonumber \\
& - & \frac{4\sqrt{2} \lambda v_\sigma}{3}\left(\frac{c_\Theta s_\Omega}{M_{\overline{\Theta}}^2 M_\Omega^2} - \frac{s_\Theta c_\Omega}{M_\Theta^2 M_{\overline{\Omega}}^2}\right)\, G_{AB}^*G_{CD}\,\epsilon^{\alpha \beta \gamma}\,\overline{D}_a\, d^{C \dagger}_{\alpha A}\,C^{-1}\,{d^C_{\beta B}}^*\, e^{C T}_C\,C^{-1}\,{q^a_{\gamma D}}\, \nonumber \\
& + & \frac{8\sqrt{2} \lambda v_\sigma}{3} \left(\frac{c_\Theta c_\Delta}{M_{\overline{\Theta}}^2 M_\Delta^2} + \frac{s_\Theta s_\Delta}{M_{\Theta}^2 M{^2_{\overline{\Delta}}}}\right)\, G_{AB}^*G_{CD}^*\,\epsilon^{\alpha \beta \gamma}\,D^a\,d^{C \dagger}_{\alpha A}\,C^{-1}\,{d^C_{\beta B}}^*\, l^\dagger_{a C}\,C^{-1}\,{d^C_{\gamma D}}^*\, \nonumber \\
& - & \frac{8 \lambda v_\sigma}{3} \left(\frac{c_{\mathbb{T}} s_\Delta}{M_{\overline{\Delta}}^2 M_{\mathbb{T}}^2}- \frac{s_{\mathbb{T}} c_\Delta}{M_\Delta^2 M_{\overline{\mathbb{T}}}^2} \right)\, G_{AB}^*G_{CD}\,\epsilon^{\alpha \beta \gamma}\, D^a\,q^{\dagger}_{\alpha b A}\,C^{-1}\,{q^*_{\beta a B}}\, l^{bT}_C\,C^{-1}\,d^{C}_{\gamma D}\, \nonumber \\
&+& {\rm h.c.}\,,\eeqa
where $c_\chi = \cos \theta_\chi$,  $s_\chi = \sin \theta_\chi$ and $\theta_\chi$ is the angle denoting the mixing between $\chi$-$\overline{\chi}$ fields analogous to the one defined earlier in Eqs. (\ref{T_redf},\ref{mixing_angle}). The operator in the first line in Eq. (\ref{op_120_0_d7}) can be identified with $\tilde{O}_1$, second with $\tilde{O}_2$, third with $\tilde{O}_5$, fourth with $\tilde{O}_6$ and the last with operator $\tilde{O}_4$ as listed in \cite{Babu:2012vb}. Note that the operator $\tilde{O}_3$ does not arise as coupling of $\mathbb{T}$ with pair of quark doublets are forbidden by flavour anti-symmetry of $G$.

Using the Fierz rearrangement, Eq. (\ref{fierz}), the operator in the second line of Eq. (\ref{op_120_0_d7}) can be expressed in terms of the one in the first line. Once the electroweak doublets acquire VEVs, the four-fermion operator arising from Eq. (\ref{op_120_0_d7}) in the physical basis can be parametrized as the following:
\beqa \label{op_120_d7}
{\cal L}^{\bf 120}_{\rm eff} &=&\tilde{g}[u^C_A,d^C_A,\nu_C,d^C_D]\,\epsilon^{\alpha \beta \gamma}\, u^{C T}_{\alpha A}\,C^{-1}\,{d^C_{\beta B}}\,\nu_C^T\,C^{-1}\, d^{C}_{\gamma D}\, \nonumber \\
& + & \tilde{g}^\prime[d^C_A,d^C_B,e^C_C,d_D]\,\epsilon^{\alpha \beta \gamma}\, d^{C \dagger}_{\alpha A}\,C^{-1}\,d^{C *}_{\beta B}\, e^{C T}_C\,C^{-1}\,d_{\gamma D}\, \nonumber \\
& + & \tilde{g}[d^C_A,d^C_B,e_C,d^C_D]\,\epsilon^{\alpha \beta \gamma}\, d^{C T}_{\alpha A}\,C^{-1}\,d^C_{\beta B}\, e^T_C\,C^{-1}\,d^C_{\gamma D}\, \nonumber \\
& + & \tilde{g}^\prime[u_A,d_B,\nu_C,d^C_D]\,\epsilon^{\alpha \beta \gamma}\, u^{\dagger}_{\alpha A}\,C^{-1}\,{d^*_{\beta B}}\, \nu^{T}_C\,C^{-1}\,d^{C}_{\gamma D}\, \nonumber \\
& + & \tilde{g}^\prime[d_A,d_B,e_C,d^C_D]\,\epsilon^{\alpha \beta \gamma}\, d^{\dagger}_{\alpha A}\,C^{-1}\,{d^*_{\beta B}}\, e^{T}_C\,C^{-1}\,d^{C}_{\gamma D}\,+\, {\rm h.c.}\,,\eeqa
with
\beqa \label{coff_op_120_d7}
 \tilde{g}[u^C_A,d^C_B,\nu_C,d^C_D] &=&\frac{8 \lambda v_\sigma v_{\overline{D}} }{3}\left[\left( \frac{1}{M{^2_{\Delta}}\,M{^2_{\overline{T}_1}}}+\frac{c_\Delta c_T}{M_\Delta^2 M_{\overline{T}_2}^2}\right) + \frac{s_\Delta s_T}{M_{\overline{\Delta}}^2 M_{T}^2}\right]
 \left(U_{u^C}^T G U_{d^C} \right)_{AB}\,\left(U_\nu^T G U_{d^C}\right)_{CD} \nonumber \\
 & + &   \frac{8\sqrt{2} \lambda v_\sigma v_{\overline{D}} }{3}\left(\frac{c_\Theta c_\Omega}{M_{\overline{\Theta}}^2 M_\Omega^2} + \frac{s_\Theta s_\Omega}{M_\Theta^2 M_{\overline{\Omega}}^2} \right)\, \left(U{_{\nu}^T} G U_{u^C} \right)_{CA}\,\left(U{_{d^C}^T} G U_{d^C} \right)_{BD}\,,\nonumber\\
\tilde{g}^\prime[d^C_A,d^C_B,e^C_C,d_D] & = &- \frac{4\sqrt{2} \lambda v_\sigma v_{\overline{D}} }{3}\left(\frac{c_\Theta s_\Omega}{M_{\overline{\Theta}}^2 M_\Omega^2}  - \frac{s_\Theta c_\Omega}{M_\Theta^2 M_{\overline{\Omega}}^2}  \right)\,\left(U_{d^C}^\dagger G^* U_{d^C}^*\right)_{AB} \,\left(U_{e^C}^T G U_{d}\right)_{CD} \,, \nonumber \\
\tilde{g}[d^C_A,d^C_B,e_C,d^C_D]& = & \frac{8\sqrt{2} \lambda v_\sigma v_D}{3} \left(\frac{c_\Theta c_\Delta}{M_{\overline{\Theta}}^2 M_\Delta^2} + \frac{s_\Theta s_\Delta}{M_{\Theta}^2 M{^2_{\overline{\Delta}}}} \right)\,\left(U_{d^C}^T G U_{d^C} \right)_{AB}\,\left(U{_ e^T} G U_{d^C}\right)_{CD}\,, \nonumber \\ 
\tilde{g}^\prime[u_A,d_B,\nu_C,d^C_D]& =& -\frac{8 \lambda v_\sigma v_D}{3} \left(\frac{c_{\mathbb{T}} s_\Delta}{M_{\overline{\Delta}}^2 M_{\mathbb{T}}^2} - \frac{s_{\mathbb{T}} c_\Delta}{M_\Delta^2 M_{\overline{\mathbb{T}}}^2} \right)\, \left(U_u^\dagger G^* U_d^* \right)_{AB}\, \left(U_\nu^T G U_{d^C} \right)_{CD} \,,\nonumber \\ 
\tilde{g}^\prime[d_A,d_B,e_C,d^C_D] & = & -\frac{8 \lambda v_\sigma v_D}{3} \left(\frac{c_{\mathbb{T}} s_\Delta}{M_{\overline{\Delta}}^2 M_{\mathbb{T}}^2} - \frac{s_{\mathbb{T}} c_\Delta}{M_\Delta^2 M_{\overline{\mathbb{T}}}^2} \right)\, \left(U_d^\dagger G^* U_d^* \right)_{AB}\, \left(U_e^T G U_{d^C} \right)_{CD}\,. \eeqa
As in the earlier case, primed and unprimed coefficients are defined in such a way that all the $\tilde{g}^\prime = 0$ there is no mixing between the different scalars and their conjugate partners.

In summary, the operators listed in Eqs. (\ref{op_126_d7},\ref{op_120_d7}) quantify the $B$, $L$ and $B-L$ violating baryon decays mediated by scalars residing in $\overline{\bf 126}_H$ and ${\bf 120}_H$, respectively, at the leading order. Like in the previous section, we have considered the possibility of the mixing between the various scalars arising from the given representation. It can be noted that the masses of $T$, $\overline{T}$, $\mathbb{T}$ and $\overline{\mathbb{T}}$ are already constrained by the leading order $d=6$ operators considered in the previous section. The $B-L$ violating decays of nucleon can, therefore, provide lower bounds on the masses of $\Delta$, $\overline{\Delta}$ residing in $\overline{\bf 126}_H$ and $\Delta$, $\overline{\Delta}$, $\Theta$, $\overline{\Theta}$, $\Omega$ and $\overline{\Omega}$ belonging to ${\bf 120}_H$ depending on their couplings with quarks and leptons and the scale of $B-L$ breaking.

\section{Nucleon decay partial widths}
\label{sec:decay_widths}
In this section, we give explicit expressions of the proton decay widths in various channels evaluated from the derived effective operators.

\subsection{$B-L$ conserving decays}
The $d=6$ operators listed previously in Eqs. (\ref{op_10},\ref{op_126},\ref{op_120}) can be parametrized in terms of the following six independent operators:
\beqa \label{gen_op}
{\cal L}_{\rm eff} & = & y[u_A,d_B,e^C_C,u^C_D]\,\, \epsilon^{\alpha \beta \gamma}\, u_{\alpha A}^T\,C^{-1}\,d_{\beta B}\, {e^C_C}^\dagger\,C^{-1}\,{u^C_{\gamma D}}^* \nonumber \\
& + & y[u^C_A,d^C_B,e_C,u_D]\,\, \epsilon^{\alpha \beta \gamma}\, u{^{CT}_{\alpha A}}\,C^{-1}\,d_{\beta B}^C\, e_C^\dagger\,C^{-1}\,u_{\gamma D}^*\, \nonumber \\
& + & y[u^C_A,d^C_B,\nu_C,d_D]\,\, \epsilon^{\alpha \beta \gamma}\, u{^{CT}_{\alpha A}}\,C^{-1}\,d_{\beta B}^C\, \nu_C^\dagger\,C^{-1}\,d_{\gamma D}^*\, \nonumber\\
& + & y^\prime[u^C_A,d^C_B,e^C_C,u^C_D]\,\, \epsilon^{\alpha \beta \gamma}\, u{^{CT}_{\alpha A}}\,C^{-1}\, d^C_{\beta B}\, e{^{CT}_C}\,C^{-1}\,u^C_{\gamma D} \nonumber \\
& + & y^\prime[u_A,d_B,e_C,u_D]\,\, \epsilon^{\alpha \beta \gamma}\, u_{\alpha A}^T\,C^{-1}\,d_{\beta B}\, e_C^T\,C^{-1}\,u_{\gamma D}\, \nonumber \\
& + & y^\prime[u_A,d_B,\nu_C,d_D]\,\, \epsilon^{\alpha \beta \gamma}\, u_{\alpha A}^T\,C^{-1}\,d_{\beta B}\, \nu_C^T\,C^{-1}\,d_{\gamma D}\,\, +{\rm h.c.}\,,\eeqa
where $y = h$, $f$ or $g$ if a single GUT scalar representation is considered. $y$ can also be a linear combination of $h$, $f$ and $g$ when more than one scalar fields are considered as discussed in the previous section. These operators match with the most general dimension six operators derived from effective theory \cite{PhysRevLett.43.1566,PhysRevLett.43.1571} but now the coefficient can be explicitly computed in terms of fundamental Yukawa couplings of a given GUT model.

To write the above operators in the usual left- and right-chiral fields, we use 
\be \label{lrnotation}
\psi = \psi_L\,,~~\psi^C = C\, \psi_R^*\,, \ee
where $C = -i \sigma^2$ in Weyl basis. This leads to
\be \label{identity}
\psi^T\,C^{-1}\,\chi = \overline{(\psi_L)^C}\,\chi_L\,,~~{\psi^C}^\dagger\,C^{-1}\,{\chi^C}^* = \overline{(\psi_R)^C}\,\chi_R\,,\ee
where $\overline{\psi^C} \equiv \psi^T C^{-1}$. Using the above identities and $\overline{\psi^C}\, \chi = \overline{\chi^C}\,\psi$, the baryon number violating operators  listed in Eq. (\ref{gen_op}) can be brought into the following convenient form:
\beqa \label{gen_op_2}
{\cal L}_{\rm eff} & = & y[u_A,d_B,e^C_C,u^C_D]\,\, \epsilon^{\alpha \beta \gamma}\, \overline{(d_{\beta B L})^C}\, u_{\alpha A L}\,\, \overline{(u_{\gamma D R})^C}\, e_{C R} \nonumber \\
& + & y^*[u^C_A,d^C_B,e_C,u_D]\,\, \epsilon^{\alpha \beta \gamma}\, \overline{(d_{\beta B R})^C}\, u_{\alpha A R}\,\, \overline{(u_{\gamma D L})^C}\, e_{C L} \nonumber \\
& + & y^*[u^C_A,d^C_B,\nu_C,d_D]\,\, \epsilon^{\alpha \beta \gamma}\,\overline{(d_{\beta B R})^C}\, u_{\alpha A R}\,\, \overline{(d_{\gamma D L})^C}\, \nu_{C L} \nonumber \\
& + & y^{\prime *}[u^C_A,d^C_B,e^C_C,u^C_D]\,\, \epsilon^{\alpha \beta \gamma}\, \overline{(d_{\beta B R})^C}\, u_{\alpha A R}\,\, \overline{(u_{\gamma D R})^C}\, e_{C R} \nonumber \\
& + & y^\prime[u_A,d_B,e_C,u_D]\,\, \epsilon^{\alpha \beta \gamma}\, \overline{(d_{\beta B L})^C}\, u_{\alpha A L}\,\, \overline{(u_{\gamma D L})^C}\, e_{C L} \nonumber \\
& + & y^\prime[u_A,d_B,\nu_C,d_D]\,\, \epsilon^{\alpha \beta \gamma}\,\overline{(d_{\beta B L})^C}\, u_{\alpha A L}\,\, \overline{(d_{\gamma D L})^C}\, \nu_{C L} +{\rm h.c.}\,.\eeqa
where $y^*[A,B,C,D] = \left(y[A,B,C,D]\right)^*$ and the same for $y^\prime$.

All the operators listed in Eq. (\ref{gen_op_2}) violate both $B$ and $L$ but conserve $B-L$ leading to decays of baryon into meson and anti-lepton. The hadronic matrix element between the baryon and meson states can be computed using the chiral perturbation theory \cite{Claudson:1981gh,Chadha:1983sj}. Using the results from \cite{JLQCD:1999dld}, one finds the following expressions for the partial decay widths for the proton decaying into various mesons \cite{Nath:2006ut}:
\beqa \label{decay_width}
%\begin{split}
\Gamma[p \to e_i^+\pi^0] &=& \frac{(m_p^2 - m_{\pi^0}^2)^2}{32\, \pi\, m_p^3 f_\pi^2} A^2 \left( \frac{1+\tilde{D}+\tilde{F}}{\sqrt{2}}\right)^2  \Big(\left| \alpha\, y[u_1,d_1,e^C_i,u^C_1] + \beta\, y^{\prime *}[u^C_1,d^C_1,e^C_i,u^C_1] \right|^2 \Big. \nonumber \\
& + & \Big. \left| \alpha\, y^*[u^C_1,d^C_1,e_i,u_1] + \beta\, y^\prime[u_1,d_1,e_i,u_1]\right|^2 \Big), \nonumber\\
\Gamma[p \to \overline{\nu}\pi^+] &=& \frac{(m_p^2 - m_{\pi^\pm}^2)^2}{32\, \pi\, m_p^3 f_\pi^2} A^2 \left(1+\tilde{D}+\tilde{F} \right)^2 \sum_{i=1}^3 \left| \alpha\, y^*[u^C_1,d^C_1,\nu_i, d_1] + \beta\, y^\prime[u_1,d_1,\nu_i, d_1] \right|^2\,, \nonumber\\
\Gamma[p \to e_i^+K^0] &=& \frac{(m_p^2 - m_{K^0}^2)^2}{32\, \pi\, m_p^3 f_\pi^2} A^2\, \frac{1}{2} \Big[\Big|C_{Li}^- - C_{Ri}^- +\frac{m_p}{m_B}(\tilde{D}-\tilde{F})\left(C_{Li}^+ - C_{Ri}^+\right)\Big|^2 \Big. \nonumber \\
& + & \Big. \Big|C_{Li}^- + C_{Ri}^- +\frac{m_p}{m_B}(\tilde{D}-\tilde{F})\left(C_{Li}^+ + C_{Ri}^+\right)\Big|^2 \Big]\,, \nonumber\\
\Gamma[p \to \overline{\nu}K^+] &=& \frac{(m_p^2 - m_{K^\pm}^2)^2}{32\, \pi\, m_p^3 f_\pi^2}\, A^2\, \sum_{i=1}^3 \Big| \frac{2\tilde{D}}{3} \frac{m_p}{m_B} C^\nu_{L1i} + \left(1+\frac{\tilde{D}+3\tilde{F}}{3} \frac{m_p}{m_B}\right) C^\nu_{L2i} \Big|^2, \nonumber\\
\Gamma[p \to e_i^+\eta] &=& \frac{(m_p^2 - m_{\eta}^2)^2}{32\, \pi\, m_p^3 f_\pi^2}\, A^2\, \frac{1}{6}\Big[ \Big|C_{L i}^+ (1-\tilde{D}+3\tilde{F}) - 2 C_{L i}^-\Big|^2 \, \nonumber\\
& + & \Big. \Big| C_{R i}^+ (1-\tilde{D}+3\tilde{F}) - 2 C_{R i}^-\Big|^2\Big]\, \Big.,\eeqa
where
\beqa \label{C_dw}
C^\pm_{L i} &=& \alpha\, y^*[u^C_1,d^C_2,e_i,u_1] \pm \beta\, y^\prime[u_1,d_2,e_i,u_1]\,,
\nonumber\\
C^\pm_{R i} &=& \alpha\, y[u_1,d_2,e^C_i,u^C_1] \pm \beta\, y^{\prime *}[u^C_1,d^C_2,e^C_i,u^C_1]\,,
\nonumber\\
C^\nu_{L1i} &=& \alpha\, y^*[u_1^C,d_2^C,\nu_i,d_1] + \beta\, y^\prime[u_1,d_2,\nu_i,d_1]\,,\nonumber \\
C^\nu_{L2i} &=& \alpha\, y^*[u_1^C,d_1^C,\nu_i,d_2] + \beta\, y^\prime[u_1,d_1,\nu_i,d_2]\,. \eeqa
Here, $m_H$ denotes the mass of hadron $H$ ($H=p,\pi^0,\pi^\pm,K^0,K^\pm,\eta$), $m_B$ is average baryon mass and $f_\pi$ is pion decay constant. $\alpha$, $\beta$, $\tilde{D}$ and $\tilde{F}$ are the parameters of the chiral Lagrangian. The factor $A$ accounts for the renormalization effects in hadronic matrix elements from the weak scale to the $m_p$.

It can be noticed from Eqs. (\ref{decay_width},\ref{C_dw}) that the relevant couplings for the proton decay into the charged leptons are $y[u_1,d_i,e^C_j,u^C_1]$, $y[u^C_1,d^C_i,e_j,u_1]$, $y^\prime[u^C_1,d^C_i,e^C_j,u^C_1]$, $y[u_1,d_i,e_j,u_1]$ and for the proton decay into the neutrinos are $y[u_1,d_i,\nu_j,d_k]$ and $y[u^C_1,d^C_i,\nu_j,d_k]$ . Considering this, the tree-level contribution mediated by ${\cal T}$, $\overline{\cal T}$ (see Eq. (\ref{coff_op_120})) vanishes due to anti-symmetric $G$ as
\be \label{cancel_G}
\left(U_{u^C}^T G U_{u^C} \right)_{11} = 0\,.\ee
These fields can induce proton-decay through dimension-6 operators which arise at loop level. The same result has been found earlier in the context of $SU(5)$ GUTs in \cite{Dorsner:2012nq} and the 1-loop diagrams which arise through an additional $W$-boson exchange have also been evaluated.  Therefore, at tree-level through dimension-6 operators, the proton decays are mediated by only $T$, $\overline{T}$ and $\mathbb{T}$, $\overline{\mathbb{T}}$ in the models with ${\bf 120}_H$.

\subsection{$B-L$ non-conserving decays}
The $B-L$ non-conserving decays of nucleons arise from the operators derived in Eqs. (\ref{op_126_d7},\ref{op_120_d7}). They can be further generalized as 
\beqa \label{gen_op_d7}
{\cal L}_{\rm eff} &=& \tilde{y}[u^C_A,d^C_A,\nu_C,d^C_D]\,\epsilon^{\alpha \beta \gamma}\, u^{C T}_{\alpha A}\,C^{-1}\,{d^C_{\beta B}}\,\nu_C^T\,C^{-1}\, d^{C}_{\gamma D}\, \nonumber \\
& + & \tilde{y}^\prime[u_A,d_B,\nu_C,d^C_D]\,\epsilon^{\alpha \beta \gamma}\, u^{\dagger}_{\alpha A}\,C^{-1}\,{d^*_{\beta B}}\, \nu^{T}_C\,C^{-1}\,d^{C}_{\gamma D}\, \nonumber \\
& + & \tilde{y}[d^C_A,d^C_B,e_C,d^C_D]\,\epsilon^{\alpha \beta \gamma}\, d^{C T}_{\alpha A}\,C^{-1}\,d^C_{\beta B}\, e^T_C\,C^{-1}\,d^C_{\gamma D}\, \nonumber \\
& + & \tilde{y}^\prime[d_A,d_B,e_C,d^C_D]\,\epsilon^{\alpha \beta \gamma}\, d^{\dagger}_{\alpha A}\,C^{-1}\,{d^*_{\beta B}}\, e^{T}_C\,C^{-1}\,d^{C}_{\gamma D}\, \nonumber \\
& + & \tilde{y}^\prime[d^C_A,d^C_B,e^C_C,d_D]\,\epsilon^{\alpha \beta \gamma}\, d^{C \dagger}_{\alpha A}\,C^{-1}\,d^{C *}_{\beta B}\, e^{C T}_C\,C^{-1}\,d_{\gamma D}\, +\, {\rm h.c.}\,,\eeqa
where $y = f$ or $g$. The above operators can be rewritten in a usual left- and right-chiral fields using identities given in Eq. (\ref{identity}) and 
\be \label{identity_2}
\psi^T C^{-1} \chi^C = \left(\overline{\psi_L}\, \chi_R \right)^*\,~~\psi^{CT} C^{-1} \chi = \overline{\psi_R}\, \chi_L \,.\ee
The above identities are obtained from the definitions given in Eq. (\ref{lrnotation}). In the new notation, we find 
\beqa \label{gen_op_2_d7}
{\cal L}_{\rm eff} &=& \tilde{y}^*[u^C_A,d^C_A,\nu_C,d^C_D]\,\epsilon^{\alpha \beta \gamma}\, \overline{(d_{\beta B R})^C}\,u_{\alpha A R}\,\, \overline{d_{\gamma D R}}\,\nu_{C L}\, \nonumber \\
& + & \tilde{y}^{\prime *}[u_A,d_B,\nu_C,d^C_D]\,\epsilon^{\alpha \beta \gamma}\,  \overline{(d_{\beta B L})^C}\,u_{\alpha A L}\,\, \overline{d_{\gamma D R}}\,\nu_{C L}\, \nonumber \\
& + & \tilde{y}^*[d^C_A,d^C_B,e_C,d^C_D]\,\epsilon^{\alpha \beta \gamma}\, \overline{(d_{\beta B R})^C}\,d_{\alpha A R}\,\, \overline{d_{\gamma D R}}\,e_{C L}\, \nonumber \\
& + & \tilde{y}^{\prime *}[d_A,d_B,e_C,d^C_D]\,\epsilon^{\alpha \beta \gamma}\, \overline{(d_{\beta B L})^C}\,d_{\alpha A L}\,\, \overline{d_{\gamma D R}}\,e_{C L}\, \nonumber \\
& + & \tilde{y}^\prime[d^C_A,d^C_B,e^C_C,d_D]\,\epsilon^{\alpha \beta \gamma}\, \overline{(d_{\beta B R})^C}\,d_{\alpha A R}\,\, \overline{d_{\gamma D L}}\,e_{C R}\, +\, {\rm h.c.}\,.\eeqa

The operators obtained above violate $B-L$ by two units and lead to processes in which a nucleon decays into lepton and a meson. The first two operators in Eq. (\ref{gen_op_2_d7}) contribute in the proton decay through channels $p \to \nu\, \pi^+$ and $p \to \nu\, K^+$. 
\beqa \label{decay_width_d7_1}
\Gamma[p \to \nu \pi^+] &=& \frac{(m_p^2 - m_{\pi^+}^2)^2}{32\, \pi\, m_p^3 f_\pi^2} A^2 \left(1+\tilde{D}+\tilde{F} \right)^2 \sum_{i=1}^3 \left| \alpha\, \tilde{y}^*[u^C_1,d^C_1,\nu_i, d^C_1] + \beta\, \tilde{y}^{\prime *}[u_1,d_1,\nu_i, d^C_1] \right|^2\,, \nonumber\\
\Gamma[p \to \nu K^+] &=& \frac{(m_p^2 - m_{K^+}^2)^2}{32\, \pi\, m_p^3 f_\pi^2}\, A^2\, \sum_{i=1}^3 \Big| \frac{2\tilde{D}}{3} \frac{m_p}{m_B} \tilde{C}^\nu_{L1i} + \left(1+\frac{\tilde{D}+3\tilde{F}}{3} \frac{m_p}{m_B}\right) \tilde{C}^\nu_{L2i} \Big|^2 \,,\eeqa 
where
\beqa \label{C_dw_d7}
\tilde{C}^\nu_{L1i} &=& \alpha\, \tilde{y}^*[u_1^C,d_2^C,\nu_i,d^C_1] + \beta\, \tilde{y}^{\prime *}[u_1,d_2,\nu_i,d^C_1]\,,\nonumber \\
\tilde{C}^\nu_{L2i} &=& \alpha\, \tilde{y}^*[u_1^C,d_1^C,\nu_i,d^C_2] + \beta\, \tilde{y}^{\prime *}[u_1,d_1,\nu_i,d^C_2]\,. \eeqa
Experimentally, the above decay modes are indistinguishable from $p \to \overline{\nu}\,\pi^+/K^+$.

The remaining three operators in Eq. (\ref{gen_op_2_d7}) do not contribute in the proton decay but induces the $B$ and $B-L$ violating decays of the neutron, for example $n \to e^-\,\pi^+$. The decay width for this can be estimated as
\beqa \label{decay_width_d7_2}
\Gamma[n \to e_i^-\pi^+] &=& \frac{(m_n^2 - m_{\pi^+}^2)^2}{32\, \pi\, m_n^3 f_\pi^2} A^2 \left( \frac{1+\tilde{D}+\tilde{F}}{\sqrt{2}}\right)^2  \Big(\left| \beta\, \tilde{y}^{\prime *}[d^C_1,d^C_1,e^C_i,d_1] \right|^2 \Big. \nonumber \\
& + & \Big. \left| \alpha\, \tilde{y}^*[d^C_1,d^C_1,e_i,d^C_1] + \beta\, \tilde{y}^{\prime *}[d_1,d_1,e_i,d_1^C]\right|^2 \Big)\,.\eeqa
It can be noticed that for $y=g$, i.e. when the operators are induced only trough the triplets residing in ${\bf 120}_H$, one finds vanishing decay width for $n \to e_i^-\pi^+$ at the leading order due to anti-symmetric nature of $G$, see Eq. (\ref{coff_op_120_d7}).

\section{Estimation for a ${\bf 10}_H+ \overline{\bf 126}_H$ model}
\label{sec:results_model}
Using the general results obtained above, we now compute the partial decay widths of the proton in a specific $SO(10)$ model. The Yukawa sector of the model consists of a complex ${\bf 10}_H$ and $\overline{\bf 126}_H$ scalar fields. Each of these contains a pair of colour singlets and $SU(2)$ doublets with $Y=\pm 1/2$, see Table \ref{tab:scalars}. A pair of linear combinations of these doublets, namely $h_u$ and $h_d$, is assumed to remain much lighter than the GUT scale and induce the electroweak scale. The Vacuum Expectation Values (VEVs) of $h_u$ and $h_d$ also generate masses for all the charged fermions. $\overline{\bf 126}_H$ also contains an SM singlet but $B-L$ charged field, VEV of which gives rise to masses for heavy RH neutrinos. This, along with the Dirac masses generated by the VEVs of $h_{u,d}$ generates naturally suppressed mass for the SM neutrinos through the type I seesaw mechanism. The viability of this framework in reproducing the correct spectrum of fermion masses and mixing parameters has been extensively studied in several works, see \cite{Babu:1992ia,Bajc:2005zf,Joshipura:2011nn,Altarelli:2013aqa,Dueck:2013gca,Meloni:2014rga,Meloni:2016rnt,Babu:2016bmy,Ohlsson:2018qpt,Mummidi:2021anm} for example. In the most recent study \cite{Mummidi:2021anm}, several viable solutions have been obtained for this model which not only reproduces the known fermion mass spectrum but can also account for the observed baryon asymmetry through Leptogenesis.

The absence of ${\bf 120}_H$ in this model implies that the proton decay in this class of theories is mediated by only the $Y=\pm 1/3$ colour triplet scalars as derived in section \ref{sec:couplings}. Moreover, the minimal model uses a $U(1)$ Peccei-Quinn symmetry \cite{Peccei:1977hh} under which ${\bf 16}_F$ has charge $+1$ while ${\bf 10}_H$ and $\overline{\bf 126}_H$ each has charge $-2$. This allows Yukawa couplings with ${\bf 10}_H$ but forbids the ones with ${\bf 10}_H^\dagger$ leaving only two Yukawa coupling matrices making the model predictive \cite{Joshipura:2011nn}. As a consequence of Peccei-Quinn symmetry, the gauge invariant term ${\bf 10}_H^2$ is forbidden and hence the components of $T$ and $\overline{T}$ of ${\bf 10}_H$ do not mix. Therefore, we have $\theta = 0$ in Eq. (\ref{mixing_angle}) and all the $h^\prime$ vanish in Eq. (\ref{op_10}). The proton decay, in this case, is governed by only three independent operators, listed as the first three in Eq. (\ref{gen_op_2}).

In general, the interaction terms in the scalar potential which gives rise to mixing of electroweak doublets can also induce mixing between different $T$ and $\overline{T}$ residing in ${\bf 10}_H$ and $\overline{\bf 126}_H$. In this case, the coefficients of the effective operators given in Eq. (\ref{gen_op_2}) are linear combinations of corresponding $h$ and $f$. The determination of the exact combination however depends on the full scalar potential. To compute the proton lifetime in this model, we adopt a simplified approach and assume that the lightest pair of triplets is dominantly arising from either of ${\bf 10}_H$ and $\overline{\bf 126}_H$. The coefficients $y$, in this case, are either $h$ or $f$ and can be fully determined from the fundamental Yukawa coupling matrix $H$ and $F$ and from the flavour rotation matrices $U_f$. These parameters can be extracted from the fermion mass fit performed in \cite{Mummidi:2021anm}. The parameters corresponding to the best fit solution are given in \cite{Mummidi:2021anm} which we reproduce here in the Appendix \ref{app:fit} for convenience. We also list the diagonalizing matrix obtained from the various Yukawa coupling matrices.

The Yukawa coupling matrices $H$ and $F$ are obtained as
\be \label{HF}
H = \frac{1}{2\sqrt{2}\,\alpha_1}\,H^\prime\,~~~F = \frac{\sqrt{3}}{4\sqrt{2}\,\alpha_2}\,F^\prime\,,
\ee
where $\alpha_{1,2}$ are factors that quantify the mixing of electroweak doublets with a constraint $|\alpha_1|^2 + |\alpha_2|^2 \le 1$, see \cite{Mummidi:2021anm} for the details. $H^\prime$ and $F^\prime$ can be determined from the fermion mass spectrum and an example numerical solution is given in Appendix \ref{app:fit}. Note that $H$ can be chosen diagonal and real without the loss of generality while $F$ is complex symmetric in this basis. The fermion mass matrices, which are linear combinations of $H$ and $F$, remains complex symmetric. Hence, one finds $U_{f^c} = U_f^*$ for $f=u,d,e$. The coefficients $h$ and $f$ then can be determined from the values of $H^\prime$, $F^\prime$ and $U_f$ given in Appendix \ref{app:fit} for the best fit solution. For the parameters in Eq. ({\ref{decay_width}), we use $\alpha = 0.01\,{\rm GeV}^3$, $\tilde{D}=0.8$, $\tilde{F}=0.46$ \cite{JLQCD:1999dld,Cabibbo:2003cu}. Further, the average baryon mass $m_B=1.15$ GeV, the pion decay constant $f_\pi=130$ MeV and the values of various hadron masses are taken from the PDG \cite{Zyla:2020zbs}. The parameter $A=1.43$ captures running effects from $M_Z$ to $m_p$. To account for the running effects between $M_{\rm GUT}$ to $M_Z$ one needs to use the values of $H$ and $F$ extracted at $M_Z$. However, we use the values obtained from the fit carried out at the GUT scale as the change in Yukawa couplings due to running is of ${\cal O}(1)$ and does not change the results significantly \cite{Alonso:2014zka}.

\subsection{Proton decay pattern}
The results obtained for various branching ratios are given in Table \ref{tab:10_res} assuming that the lightest $T$ and $\overline{T}$ are dominantly the ones residing in ${\bf 10}_H$. Similarly, the branching ratios computed, assuming that the lightest pair of triplets originates from $\overline{\bf 126}_H$, are listed in Table \ref{tab:126_res}. Note that the branching ratios determined in Table \ref{tab:10_res} and \ref{tab:126_res} do not depend on the unknown parameter $\alpha_{1,2}$ appearing in Eq. (\ref{HF}).
%%%%%%%%%%%%%
\begin{table}[t]
\begin{center}
\begin{tabular}{lccc} 
\hline
\hline
Branching ratio [\%]& ~~~$M_T \ll M_{\overline{T}}$~~~  & ~~~$M_T \gg M_{\overline{T}}$~~~ & ~~~$M_T = M_{\overline{T}}$~~~\\
\hline
${\rm BR}[p\to e^+ \pi^0]$ & $< 1$ & $<1$ & $< 1$\\
${\rm BR}[p\to \mu^+ \pi^0]$  & $7$ & $< 1$ & $< 1$\\
${\rm BR}[p\to \bar{\nu} \pi^+]$  & $0$ & $15$ & $14$\\
${\rm BR}[p\to e^+ K^0]$ & $< 1$ & $< 1$ & $< 1$\\
${\rm BR}[p\to \mu^+ K^0 ]$  & $93$ & $<1$ & $2$\\
${\rm BR}[p\to \bar{\nu} K^+]$  & $0$ & $84$ & $83$\\
${\rm BR}[p\to e^+ \eta]$ & $<1$  &  $<1$ & $<1$\\
${\rm BR}[p\to \mu^+ \eta]$  & $<1$ &  $<1$ & $< 1$\\
\hline
\end{tabular}
\end{center}
\caption{Proton decay branching fractions estimated for the best fit solution for various hierarchies among the masses of $T$ and $\overline{T}$ residing in ${\bf 10}_H$.}
\label{tab:10_res}
\end{table}
%%%%%%%%%%%%%%%%%%%%
%%%%%%%%%%%%%
\begin{table}[t]
\begin{center}
\begin{tabular}{lccc} 
\hline
\hline
Branching ratio [\%]& ~~~$M_{T_{1,2}} \ll M_{\overline{T}}$~~~  & ~~~$M_{T_{1,2}} \gg M_{\overline{T}}$~~~ & ~~~$M_{T_{1,2}} = M_{\overline{T}}$~~~\\
\hline
${\rm BR}[p\to e^+ \pi^0]$ & $< 1$ & $<1$ & $< 1$\\
${\rm BR}[p\to \mu^+ \pi^0]$  & $11$ & $< 1$ & $< 1$\\
${\rm BR}[p\to \bar{\nu} \pi^+]$  & $0$ & $11$ & $11$\\
${\rm BR}[p\to e^+ K^0]$ & $< 1$ & $< 1$ & $< 1$\\
${\rm BR}[p\to \mu^+ K^0 ]$  & $88$ & $10$ & $11$\\
${\rm BR}[p\to \bar{\nu} K^+]$  & $0$ & $78$ & $77$\\
${\rm BR}[p\to e^+ \eta]$ & $<1$  &  $<1$ & $<1$\\
${\rm BR}[p\to \mu^+ \eta]$  & $<1$ &  $<1$ & $< 1$\\
\hline
\end{tabular}
\end{center}
\caption{Proton decay branching fractions estimated for the best fit solution for various hierarchies among the masses of $T_{1,2}$ and $\overline{T}$ residing in $\overline{{\bf 126}}_H$.}
\label{tab:126_res}
\end{table}
%%%%%%%%%%%%%%%%%%%%

The branching ratios of proton decay obtained in Table \ref{tab:10_res} and \ref{tab:126_res} are qualitatively very different from the ones computed assuming the dominant contribution from the gauge boson mediation. For transparent comparison, we also compute the latter taking into account the flavour effects for the best fit solution. The details of computation are given in Appendix \ref{app:gauge} and the results are listed in Table \ref{tab:gauge_res}. The important observations are:
\begin{enumerate}[(a)]
\item Unlike in the case of gauge boson mediation in which the proton preferably decays into $\pi^0\,e^+$ or $\pi^+ \overline{\nu}$, the scalar induced proton decay favours the channels involving $K^0\,\mu^+$ or $K^+\,\overline{\nu}$.\label{aa}
\item Typically, one finds ${\rm BR}[p\to e^+ \pi^0] \ll {\rm BR}[p\to \mu^+ \pi^0]$. This is in contrast to the gauge boson mediated decays in which one typically finds ${\rm BR}[p\to e^+ \pi^0] \gg {\rm BR}[p\to \mu^+ \pi^0]$.\label{ab} 
\item When $\overline{T}$ is lighter than $T$, the proton decays dominantly in the channels involving charged mesons and neutrinos. As $T$ does not couple to the neutrinos, proton decays mediated by it results into the neutral mesons and charged leptons.\label{ac}
\item A comparison between Table \ref{tab:10_res} and \ref{tab:126_res} suggests that the decay patterns of the proton do not significantly depend on whether the lightest $T$ and $\overline{T}$ originate dominantly from ${\bf 10}_H$ or $\overline{\bf 126}_H$. This indicates that the qualitative results would remain unaltered even if the lightest pair, $T$ and $\overline{T}$, are general linear combinations of triplets and anti-triplets, respectively, residing in ${\bf 10}_H$ and $\overline{\bf 126}_H$.\label{ad}
\end{enumerate}
To check the robustness of the above observations, we evaluate the proton decay branching ratios for several solutions, with acceptable $\chi^2$ values at the minimum, determined in \cite{Mummidi:2021anm}. The results are displayed in Figs. \ref{fig1} and \ref{fig2}.
%%%%%%%%%%%%%%%
\begin{figure}[t]
\centering
\subfigure{\includegraphics[width=0.33\textwidth]{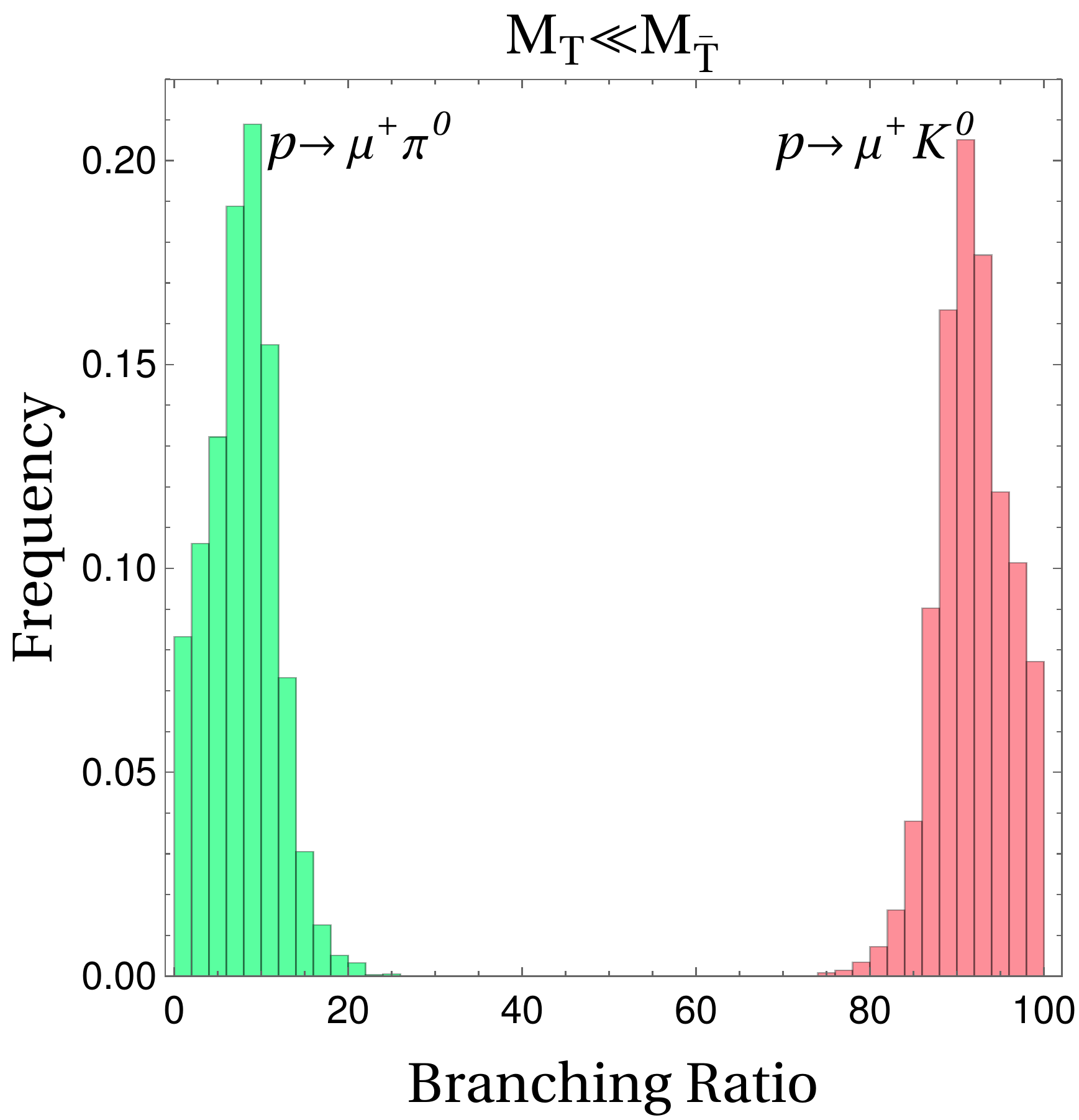}}
\subfigure{\includegraphics[width=0.325\textwidth]{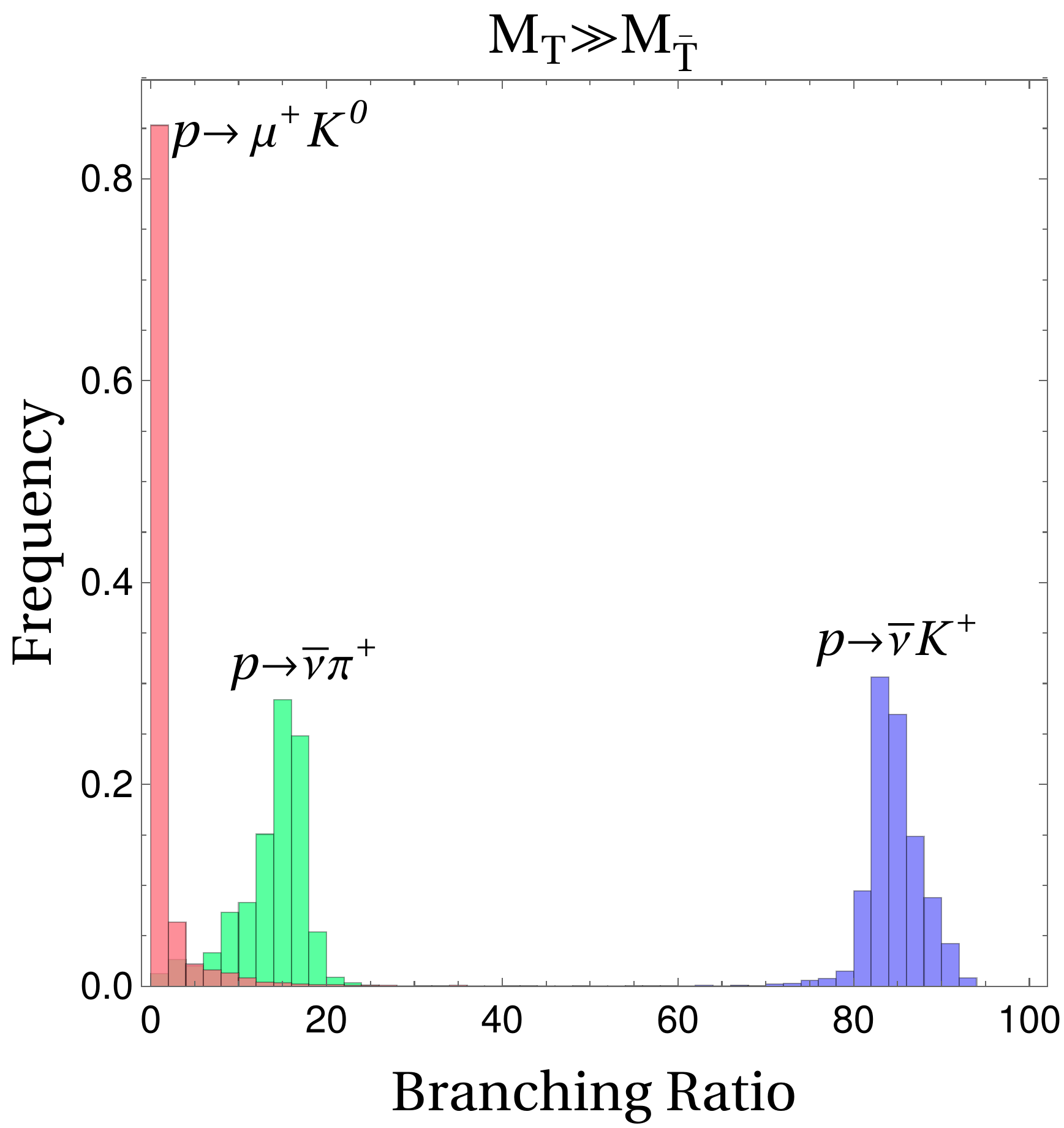}}
\subfigure{\includegraphics[width=0.325\textwidth]{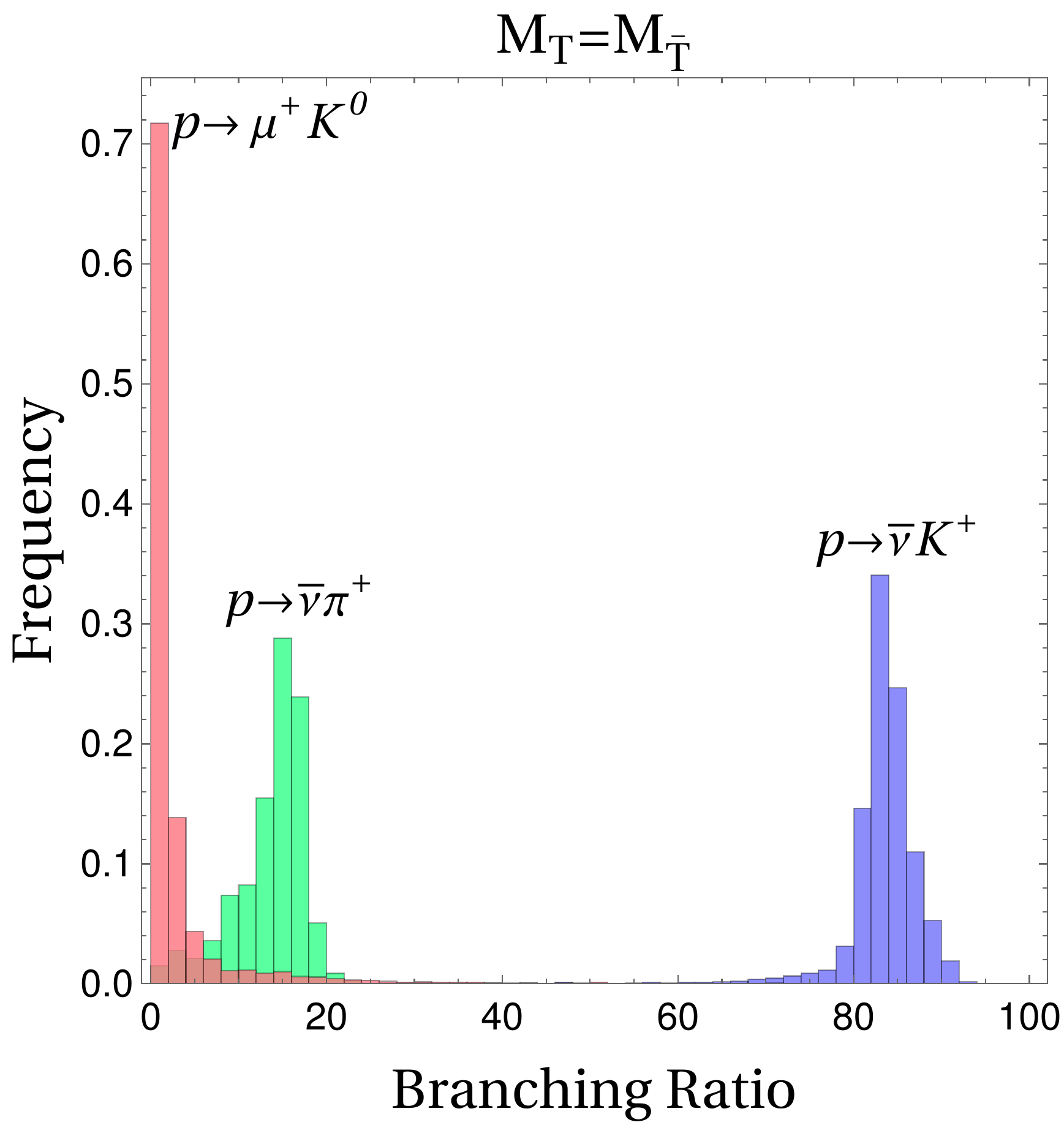}}
\caption{Scalar mediated proton decay spectrum for solutions obtained in case of the minimal renormalizable non-supersymmetric $SO(10)$ model. Various branching ratios are evaluated assuming that the lightest pair of colour triplets, $T$ and $\overline{T}$, arise dominantly from ${\bf 10}_H$.}
\label{fig1}
\end{figure}
%%%%%%%%%%%%%%
%%%%%%%%%%%%%%%
\begin{figure}
\centering
\subfigure{\includegraphics[width=0.325\textwidth]{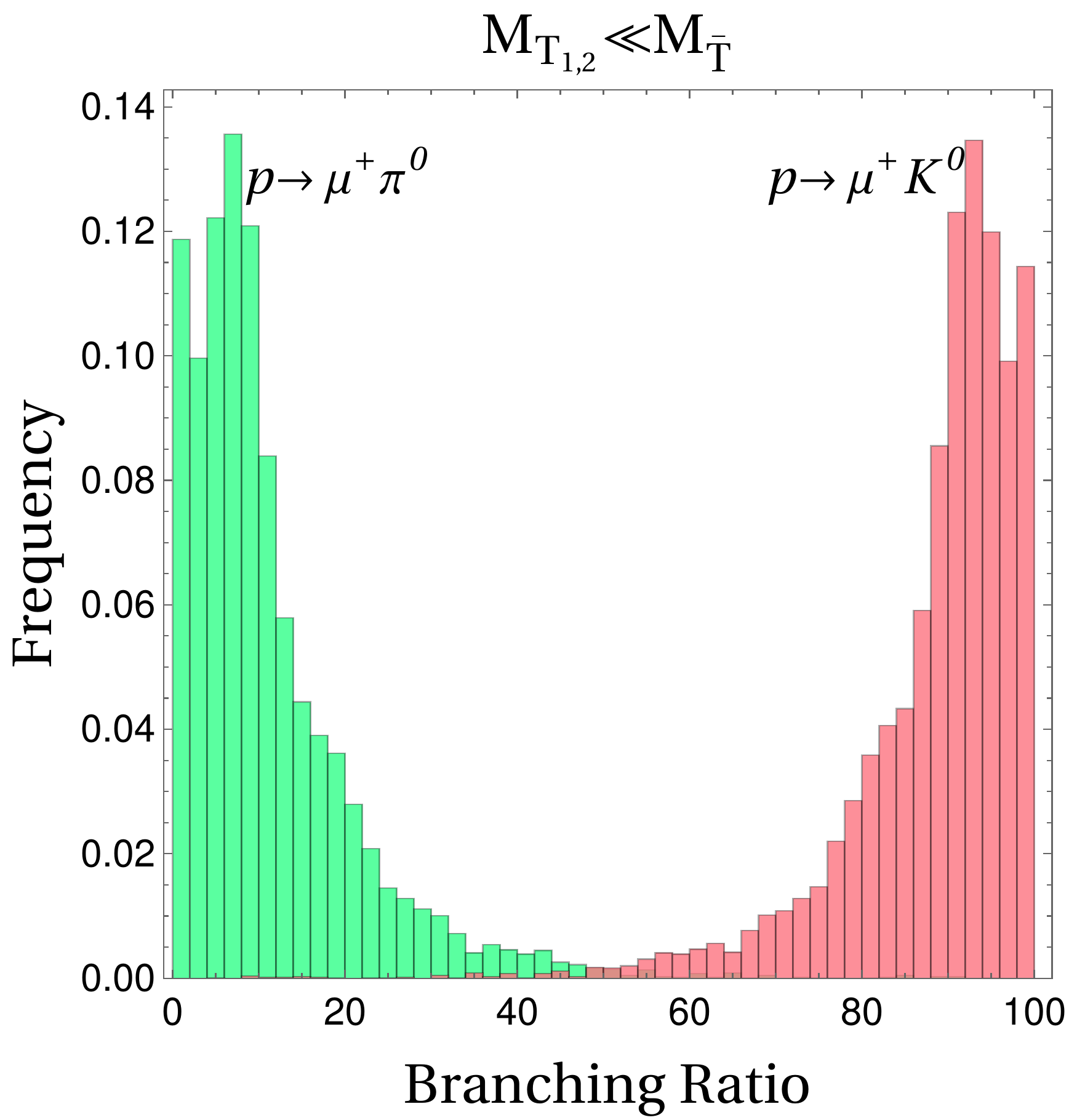}}
\subfigure{\includegraphics[width=0.325\textwidth]{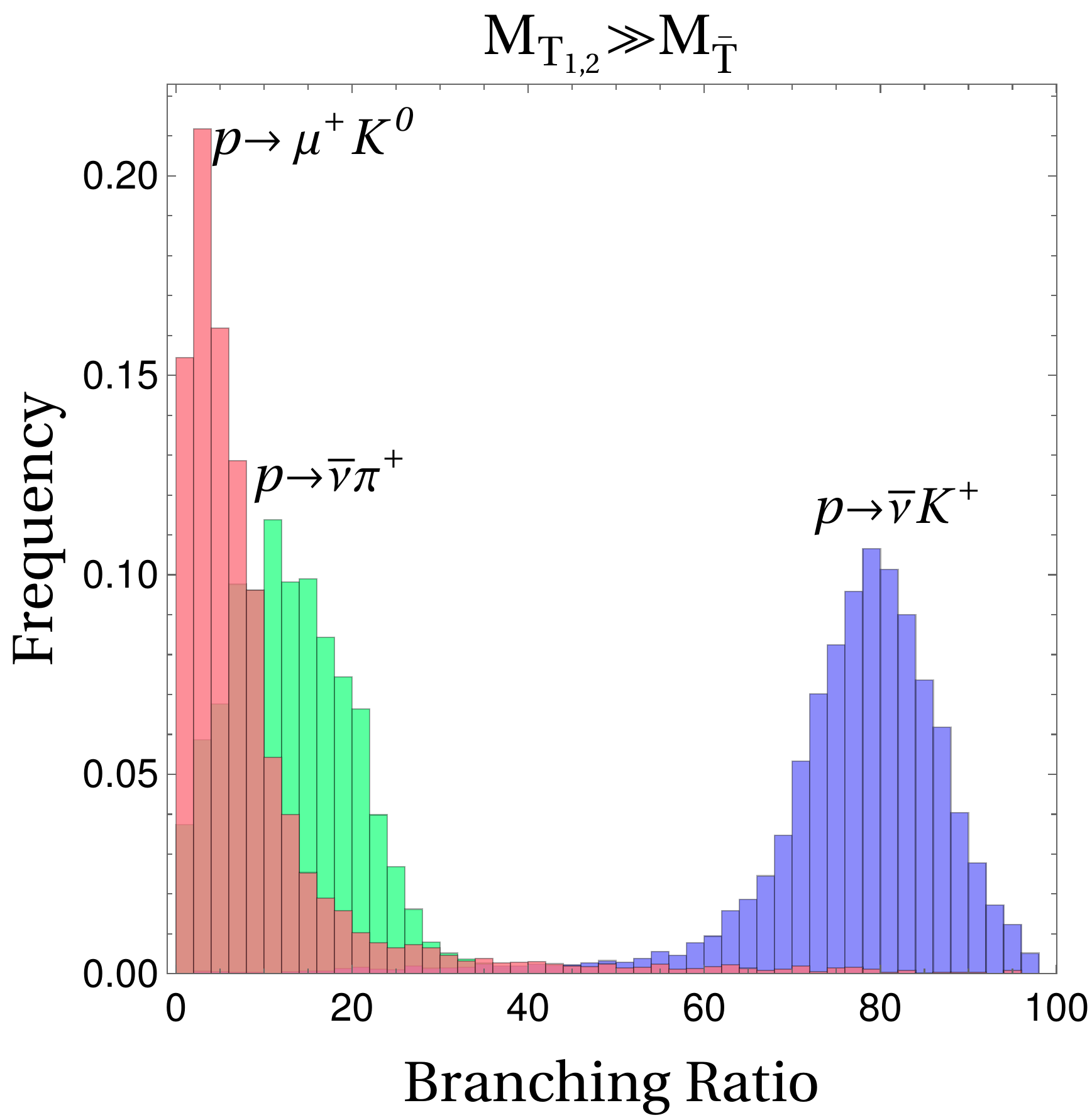}}
\subfigure{\includegraphics[width=0.325\textwidth]{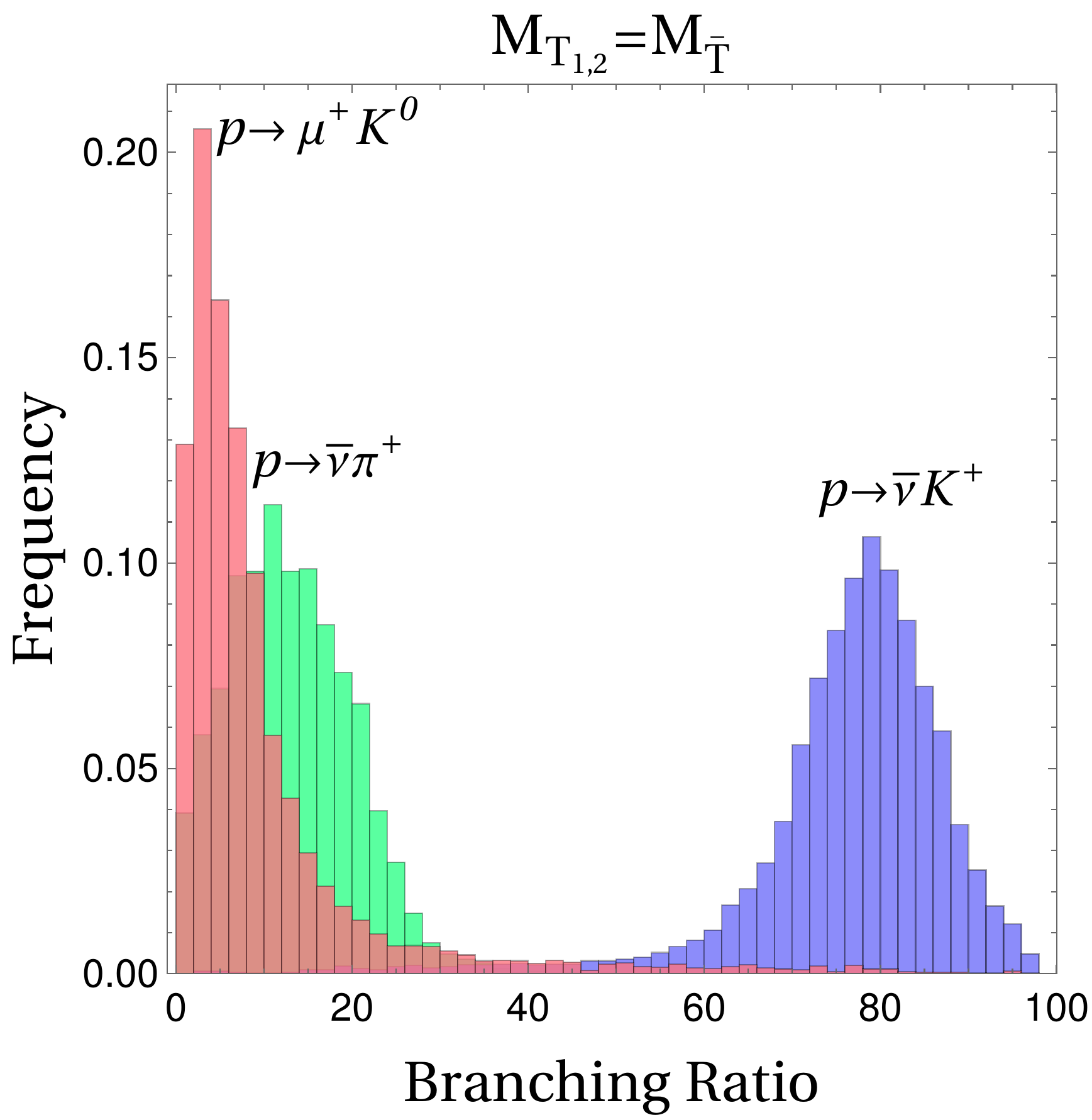}}
\caption{Same as Fig. \ref{fig1} but assuming that the lightest triplets arise dominantly from $\overline{\bf 126}_H$.}
\label{fig2}
\end{figure}
%%%%%%%%%%%%%%
It can be observed from these figures that the predictions of various branching ratios reported in Tables \ref{tab:10_res} and \ref{tab:126_res} do not change significantly even for the other viable solutions.

The noteworthy features of the proton decay spectrum listed as \ref{aa}, \ref{ab}, \ref{ac}, \ref{ad} can be understood from the flavour structure. The realistic fermion mass spectrum in the underlying model leads to the following hierarchical structures for $H$ and $F$:
\be \label{HF_form}
H \sim \frac{\lambda^4}{\alpha_1} \left(\ba{ccc} \lambda^7 & 0 & 0\\
0 & \lambda^3 & 0\\
0 & 0 & 1 \ea\right)\,,~~
F \sim \frac{\lambda^4}{\alpha_2} \left(\ba{ccc} \lambda^5 & \lambda^4 & \lambda^3\\
\lambda^4 & \lambda^3 & \lambda^2\\
\lambda^3 & \lambda^2 & \lambda \ea\right)\,,\ee
where $\lambda = 0.23$ is Cabibbo angle and we have suppressed ${\cal O}(1)$ coefficients. The Yukawa matrices of the charged fermions, $Y_{u,d,e}$, are linear combinations of $H$ and $F$. The unitary matrices which diagonalize $Y_{u,d,e}$ and the neutrino mass matrix have the following  generic form 
\be \label{U_form}
U_u \sim U_d \sim U_e \sim\left(\ba{ccc} 1 & \lambda & \lambda^3\\
\lambda & 1 & \lambda^2 \\
\lambda^3 & \lambda^2 & 1 \ea\right)\,,~~U_\nu \sim \left( \ba{ccc} &  &  \\  & {\cal O}(1) & \\  &  &  \ea \right)\,\ee
where $U_\nu$ is matrix with all the elements of ${\cal O}(1)$.  $U_{u,d,e}$ are CKM-like and lead to small quark mixing while the large mixing in PMNS matrix arise through $U_\nu$.  From Eqs. (\ref{HF_form},\ref{U_form}), we find
\beqa \label{UHU}
U_f^T\,H\,U_{f^\prime} &\sim & \frac{\lambda^4}{\alpha_1} \left(\ba{ccc} \lambda^5 & \lambda^4 & \lambda^3\\ \lambda^4 & \lambda^3 & \lambda^2\\
\lambda^3 & \lambda^2 & 1 \ea\right)\,,~~U_f^T\,F\,U_{f^\prime} \sim  \frac{\lambda^4}{\alpha_2} \left(\ba{ccc} \lambda^5 & \lambda^4 & \lambda^3\\ \lambda^4 & \lambda^3 & \lambda^2\\
\lambda^3 & \lambda^2 & \lambda \ea\right)\,, \nonumber \\
U_\nu^T\,H\,U_d &\sim & \frac{\lambda^4}{\alpha_1} \left(\ba{ccc} \lambda^3 & \lambda^2 & 1\\ \lambda^3 & \lambda^2 & 1\\
\lambda^3 & \lambda^2 & 1 \ea\right)\,,~~U_\nu^T\,F\,U_d \sim  \frac{\lambda^4}{\alpha_2} \left(\ba{ccc} \lambda^3 & \lambda^2 & \lambda\\ \lambda^3 & \lambda^2 & \lambda\\
\lambda^3 & \lambda^2 & \lambda \ea\right)\,,\eeqa
where $f,f^\prime = u,d,e$.

Substitution of the above results in Eqs. (\ref{coff_op_10},\ref{coff_op_126},\ref{decay_width}), one finds
\beqa \label{pattern_a}
\frac{\Gamma[p \to e_i^+\,\pi^0]}{\Gamma[p \to e_i^+\,K^0]} & \simeq & \frac{(m_p^2 - m{^2_{\pi^0}})^2}{(m_p^2 - m{^2_{K^0}})^2}\,\frac{(1+\tilde{D}+\tilde{F})^2}{2 \left(1+\frac{m_p}{m_B}(\tilde{D}-\tilde{F})\right)^2}\, {\lambda}^2 \simeq 3 \lambda^2\,, \nonumber \\
\frac{\Gamma[p \to \overline{\nu}\,\pi^+]}{\Gamma[p \to \overline{\nu}\,K^+]} & \simeq & \frac{(m_p^2 - m{^2_{\pi^+}})^2}{(m_p^2 - m{^2_{K^+}})^2}\,\frac{(1+\tilde{D}+\tilde{F})^2}{\left(1+\frac{m_p}{m_B}(\tilde{D}+\tilde{F})\right)^2}\,  {\lambda}^2 \simeq 2 \lambda^2\,.\eeqa
This explains the observation listed as point (a) above. Further, 
\be \label{pattern_b}
\frac{\Gamma[p \to e^+\,\pi^0]}{\Gamma[p \to \mu^+\,\pi^0]} \simeq \frac{\Gamma[p \to e^+\,K^0]}{\Gamma[p \to \mu^+\,K^0]} \simeq \lambda^2\,, \ee
leads to the result (b).  Moreover, (c) can be understood from the fact that
\be \label{}
\frac{\Gamma[p \to \mu^+\,K^0]}{\Gamma[p \to \overline{\nu}\,K^+]}  \simeq  \frac{(m_p^2 - m{^2_{K^0}})^2}{(m_p^2 - m{^2_{K^+}})^2}\,\frac{\left(1+\frac{m_p}{m_B}(\tilde{D}-\tilde{F})\right)^2}{\left(1+\frac{m_p}{m_B}(\tilde{D}+\tilde{F})\right)^2}\,  \frac{4 {\lambda}^2}{9} \simeq 0.2 \lambda^2\,.\ee
Also, it can be seen from Eq. (\ref{UHU}) that the flavour structure of couplings relevant for the proton decay are similar in case of ${\bf 10}_H$ and $\overline{\bf 126}_H$. This provides justification for the observation made in (d).

\subsection{Limits on the masses of $T$ and $\overline{T}$}
Finally, we use the current experimental limits on the lifetime of proton decay in various channels to obtain the most stringent limit on the masses of $T$ and $\overline{T}$ in the considered model. When the lightest $T$ is dominantly given by the one residing in ${\bf 10}_H$, the strongest limit on its mass comes from the decay channels involving $\mu^+$. We find from explicit computation for the best fit solution,
\beqa \label{limit_T_10}
\tau/{\rm BR}[p \to \mu^+\, K^0] &=& 1.6 \times 10^{33}\,{\rm yrs}\,\times\left(\frac{\alpha_1}{0.1}\right)^4 \times \left(\frac{M_T}{1.4 \times 10^{11}\,{\rm GeV}} \right)^4\,, \nonumber \\
\tau/{\rm BR}[p \to \mu^+\, \pi^0] &=& 1.6 \times 10^{34}\,{\rm yrs}\,\times\left(\frac{\alpha_1}{0.1}\right)^4 \times \left(\frac{M_T}{1.3 \times 10^{11}\,{\rm GeV}} \right)^4\,, \eeqa
where the first factor on the right hand side in the above equations are the current experimental lower bounds on the lifetime of proton decaying in the respective channels. Similarly for $\overline{T}$ dominantly arising from ${\bf 10}_H$, the most stringent upper bound comes from the proton decaying into neutrinos and $K^+$:
\beqa \label{limit_Tbar_10}
\tau/{\rm BR}[p \to \overline{\nu}\, K^+] &=& 5.9 \times 10^{33}\,{\rm yrs}\,\times\left(\frac{\alpha_1}{0.1}\right)^4 \times \left(\frac{M_{\overline{T}}}{6.4 \times 10^{11}\,{\rm GeV}} \right)^4\,,
\eeqa
The experimental limits on the lifetimes used for various channels in the above equations are taken from \cite{Super-Kamiokande:2020wjk,Super-Kamiokande:2012zik,Super-Kamiokande:2014otb}.

If the lightest $T$ and $\overline{T}$ arise dominantly from $\overline{\bf 126}_H$, we find 
\beqa \label{limit_T_126}
\tau/{\rm BR}[p \to \mu^+\, K^0] &=& 1.6 \times 10^{33}\,{\rm yrs}\,\times\left(\frac{\alpha_2}{0.1}\right)^4 \times \left(\frac{M_{T_i}}{2.5 \times 10^{10}\,{\rm GeV}} \right)^4\,, \nonumber \\
\tau/{\rm BR}[p \to \mu^+\, \pi^0] &=& 1.6 \times 10^{34}\,{\rm yrs}\,\times\left(\frac{\alpha_2}{0.1}\right)^4 \times \left(\frac{M_{T_i}}{2.7 \times 10^{10}\,{\rm GeV}} \right)^4\,,
\eeqa
and 
\beqa \label{limit_Tbar_126}
\tau/{\rm BR}[p \to \overline{\nu}\, K^+] &=& 5.9 \times 10^{33}\,{\rm yrs}\,\times\left(\frac{\alpha_2}{0.1}\right)^4 \times \left(\frac{M_{\overline{T}}}{1.1 \times 10^{11}\,{\rm GeV}} \right)^4\,.\eeqa
It can be noticed that the limits on the masses of $T$ and $\overline{T}$ are slightly lower than the ones given in Eqs. (\ref{limit_T_10},\ref{limit_Tbar_10}). This is due to the fact that the magnitude of Yukawa couplings with $\overline{\bf 126}_H$ are somewhat smaller than those with ${\bf 10}_H$. 

 \subsection{Limit on the mass of $\Delta$}
In addition to the $d=6$ operators discussed above, a $d=7$ operator arises within the model through $B-L$ violating mixing between $\overline{T}$ and $\Delta$ as discussed in section \ref{sec:operators_d7}. This induces decay $p \to \nu\,\pi^+$ and $p \to \nu\,K^+$. Since the mass of $\overline{T}$  is already constrained by $d=6$ operators, the lower bound on the mass of $\Delta$ can be inferred from non-observation of the proton decay. To estimate this, we take $\lambda = 1$, $\alpha_2=0.1$, $v_{\overline{D}} = 174$ GeV and $M_{\overline{T}} = 1.1 \times 10^{11}$ GeV from Eq. (\ref{limit_Tbar_126}). Substituting these values in Eqs. (\ref{coff_op_126_d7},\ref{decay_width_d7_1}), we find
\beqa \label{limit_delta}
\tau/{\rm BR}[p \to \nu\, K^+] &=& 5.9 \times 10^{33}\,{\rm yrs}\,\times\left(\frac{10^{11}\,{\rm GeV}}{v_\sigma}\right)^2 \times \left(\frac{M_{\Delta}}{7.0 \times 10^{6}\,{\rm GeV}} \right)^2\,.\eeqa
The obtained lower bound on $M_\Delta$ is two orders of magnitude smaller than the one obtained in \cite{Babu:2012vb} for the same value of $B-L$ breaking scale. The difference is due to the values of Yukawa couplings and an additional factor of $(4/15)^2$ in Eq. (\ref{coff_op_126_d7}) which arise due to Clebsch-Gordan coefficient.

\section{Summary and Discussions}
\label{sec:summary}
Proton decay is a window to peep high energy phenomena from low energy. Its observation would conclusively discard the conservation of the baryon number and strengthen the ambition to unify fundamental interactions. Moreover, it can also provide useful insight into the nature of theory in ultra-violet from which $B$ and $L$ violations originate. The latter requires the computation of nucleon decay widths in specific models identifying all possible sources. We carry out such an exercise for renormalizable $SO(10)$ GUT models.  Classifying the most general scalar spectrum of such theories, we compute explicitly the couplings of various scalars which can induce baryon and lepton number violating decays of baryons at the tree level. Effective operators are computed by integrating out the scalar fields considering also possible mixing terms between the scalars residing in the  given GUT multiplet.  We compute $d=6$ operators which conserve $B-L$ and also derive $d=7$ operators which violate $B-L$. The latter can lead to nucleon decay modes which are less studied. We then express the proton decay widths in terms of these operators and provide a comprehensive analysis of scalar mediated proton decay in a particular model. The noteworthy features of our general analysis are the following:
\begin{itemize}
\item Even though there exists several multiplets charged under $B-L$ in models with ${\bf 10}_H$, $\overline{\bf 126}_H$ and ${\bf 120}_H$, the $d=6$ operators which can induce $B$ and $L$ non-conserving (but $B-L$ conserving) baryon decays arise from only three pairs of color triplet fields: $T(3,1,-\frac{1}{3})$, ${\cal T}(3,1,-\frac{4}{3})$, $\mathbb{T}(3,3,-\frac{1}{3})$ and their conjugates.
\item In the models with ${\bf 10}_H$ and/or $\overline{\bf 126}_H$, only $T$ and $\overline{T}$ mediate the proton decay. Although $\overline{\bf 126}_H$ contains ${\cal T}$ and $\mathbb{T}$ fields, they have only lepto-quark couplings.
\item When ${\bf 120}_H$ is present in the model, the contribution to proton decay from ${\cal T}$-$\overline{\cal T}$ vanishes at tree-level due to the anti-symmetric nature of Yukawa couplings with ${\bf 120}_H$. Therefore, these fields can contribute to proton decay only at the loop level.

\item The $B-L$ non-conserving nucleon decays, which arise through $d=7$ operators at the leading order, can be mediated in general by $\Theta(3,1,\frac{2}{3})$, $\Delta(3,2,\frac{1}{6})$, $\Omega(3,2,\frac{7}{6})$ and their conjugate partners. In the models without ${\bf 120}_H$, only $\Delta$ can induce such decays.
\end{itemize}

In the context of a minimal model based on ${\bf 10}_H$ and $\overline{\bf 126}_H$ with Peccei-Quinn symmetry, we find that when scalar mediated contributions dominate, the proton decay spectrum can be quite different from the one typically anticipated. Several important aspects of the proton decay spectrum are listed in the section \ref{sec:results_model}. Proton dominantly decays into $\overline{\nu}\, K^+$ or $\mu^+\,K^0$ for lighter $\overline{T}$ or $T$, respectively. Moreover, one finds ${\rm BR}[p\to \mu^+ \pi^0] \gg {\rm BR}[p\to e^+ \pi^0]$. Both these features are distinct from gauge boson mediated proton decays in which proton preferably decays into $\overline{\nu}\, \pi^+$ or $e^+\,\pi^0$ and typically ${\rm BR}[p\to \mu^+ \pi^0] \ll {\rm BR}[p\to e^+ \pi^0]$.  These features mainly arise from the difference between the Yukawa and gauge couplings. The first is more sensitive to the flavour structure of the underlying  GUT model and hence, if scalar mediated contributions dominate, the proton decay can provide very useful insight into the Yukawa structure of the theory.

\section*{Acknowledgements}
This work is partially supported under MATRICS project (MTR/2021/000049) by the Science \& Engineering Research Board (SERB), Department of Science and Technology (DST), Government of India. 

\appendix
\section{Solution for ${\bf 10}_H+ \overline{\bf 126}_H$ model}
\label{app:fit}
In this Appendix, we give necessary parameters of the best fit solution obtained in the recent work \cite{Mummidi:2021anm}. The solution is obtained for a minimal non-supersymmetric  ${\bf 10}_H+ \overline{\bf 126}_H$ model with $U(1)$ Peccei-Quinn symmetry. One finds the best fit parameters for the Yukawa couplings as
\beqa \label{input_prm}
H^\prime &=& \left(
\begin{array}{ccc}
 0.00023 & 0 & 0 \\
 0 & -0.04811 & 0 \\
 0 & 0 & -5.79504 \\
\end{array}
\right) \times 10^{-3}\,, \nonumber \\
F^\prime &=&  \left(
\begin{array}{ccc}
 -0.0088+0.0178 i & 0.0475\, -0.0889 i & 0.4635\, +0.6797 i \\
 0.0475\, -0.0889 i & 1.1279\, +0.5108 i & -1.2218-2.5921 i \\
 0.4635\, +0.6797 i & -1.2218-2.5921 i & 5.4683\, -5.9856 i \\
\end{array}
\right)\times 10^{-4}\,. \eeqa
The above with $r =  77.4189$, $s = 0.3140 - 0.0282\, i\,$ and $v_S^\prime = 9.84 \times 10^{14}\,{\rm GeV}\,$ leads to the following effective Yukawa matrices for the quark and leptons:
\beqa \label{Eff_Yuk}
Y_d  &=& H^\prime + F^\prime \,,~~Y_u  =  r\,(H^\prime + s F^\prime)\,, \nonumber \\
Y_e  &=& H^\prime - 3 F^\prime \,, ~~Y_ \nu = r\,(H^\prime -3 s F^\prime)\,, ~~M_R  = v_S^\prime\, F^\prime \,. \eeqa
The light neutrino mass matrix is obtained as $M_\nu = - v_u^2\, Y_\nu\,M_R^{-1}\, Y_\nu^T$, 
where $v_{u,d} = \langle h_{u,d} \rangle$, $v_u^2 + v_d^2 \equiv v^2 = (174\, {\rm GeV})^2$ and $v_u/v_d \equiv \tan\beta = 1.5$. The mass matrices for the charged fermions are given by $M_{d,e} \equiv v_d\, Y_{d,e}$ and $M_u \equiv v_u Y_u$.

The unitary matrices $U_f$ and $U_{f^C}$ diagonalize $M_f$ ($f=d,u,e,\nu$) such that $U_f^\dagger M_f M_f^\dagger U_f = U_{f^C}^\dagger M_f^\dagger M_f U_{f^C} \equiv D^2_f$. Since $M_f$ are symmetric, one finds $U_f = U_{f^C}^*$. Numerically obtained $U_f$ are as the following.
\beqa \label{input_Uf}
U_d &=& U_{d^C}^* =  \left(
\begin{array}{ccc}
 -0.9497-0.2658 i & -0.0895+0.1388 i & -0.0101-0.0116 i \\
 -0.1053+0.1272 i & -0.5434-0.8213 i & 0.0276\, +0.0463 i \\
 -0.0157 & 0.0538 & 0.9984 \\
\end{array}
\right)\,, \nonumber\\ 
U_u &=& U_{u^C}^* =  \left(
\begin{array}{ccc}
 -0.4028-0.8952 i & 0.1456\, +0.123 i & -0.003-0.0034 i \\
 -0.0213-0.1895 i & -0.5203-0.8323 i & 0.0086\, +0.0135 i \\
 -0.0016 & 0.0165 & 0.9999 \\
\end{array}
\right)\,, \nonumber\\ 
U_e &=& U_{e^C}^* = \left(
\begin{array}{ccc}
 -0.3995-0.916 i & -0.0093+0.0126 i & 0.0115\, +0.03 i \\
 -0.0161-0.0074 i & -0.2511-0.9618 i & -0.0268-0.1043 i \\
 0.031 & -0.108 & 0.9937 \\
\end{array}
\right)\,, \nonumber\\ 
U_\nu & = & \left(
\begin{array}{ccc}
 -0.4683+0.689 i & 0.3113\, -0.4343 i & 0.0721\, -0.1233 i \\
 0.151\, -0.2689 i & 0.3553\, -0.5592 i & -0.357+0.5818 i \\
 0.4591 & 0.5249 & 0.7168 \\
\end{array}
\right)\,. \eeqa
The matrices $H^\prime$, $F^\prime$ and $U_f$ are used to compute the proton decay spectrum as explained in the section \ref{sec:results_model}.

\section{Proton decay spectrum from gauge boson mediations}
\label{app:gauge}
In this Appendix, we compute gauge mediated proton decay partial widths for quantitative comparison with the scalar induced contributions. The currents associated with $B-L$ charged gauge bosons, $X \sim (3,2,-5/6)$ and $Y \sim (3,2,1/6)$ with belong to adjoint representation of $SO(10)$, are given by \footnote{There also exists vector boson $Z\sim (3,1,2/3)$ with $B-L$ charge $4/3$. However, it does not induce proton  decay by itself \cite{Buchmuller:2019ipg}.}\cite{Machacek:1979tx}
\beqa \label{LXY}
-{\cal L}_{X,Y} & = & i\frac{g_{10}}{\sqrt{2}}\,\Big[\overline{X}_\mu \left(\overline{q}_A \gamma^\mu u^C_A + \overline{e^C}_A \gamma^\mu q_A - \overline{d^C}_A \gamma^\mu l_A \right) \Big. \nonumber \\
& + &  \Big. \overline{Y}_\mu \left(\overline{q}_A \gamma^\mu d^C_A + \overline{\nu^C}_A \gamma^\mu q_A - \overline{u^C}_A \gamma^\mu l_A \right) \Big]\,+\,{\rm h.c.}\,,\eeqa
where $g_{10}$ is unified gauge coupling and we have supressed the $SU(3)$ and $SU(2)$ indices for simplicity. Eliminating $X$ and $Y$ boson using classical equations of motions, the effective operators relevant for the proton decay are derived as:
\beqa \label{LXY_2}
{\cal L}_{\rm eff} & = & \frac{g_{10}^2}{2 M_X^2} \left(\overline{u^C}_A \gamma^\mu q_A\, \left(\overline{e^C}_B \gamma_\mu q_B - \overline{d^C}_B \gamma_\mu l_B\right)\right) \nonumber \\
& - & \frac{g_{10}^2}{2 M_Y^2} \left(\overline{d^C}_A \gamma^\mu q_A\, \overline{u^C}_B \gamma_\mu l_B\right)\,+\, {\rm h.c.}\,.\eeqa

Using Fierz reordering and fermion field redefinitions, $f \to U_{f} f$, we obtain (see \cite{Buchmuller:2019ipg} for details)
\beqa \label{LXY_3}
{\cal L}_{\rm eff} & = & k[u_A,d_B,e^C_C,u^C_D]\left(\overline{e^C}_C\, \overline{u^C}_D\, u_A\, d_B\right) \nonumber \\
& + & k[u^C_A,d^C_B,e_C,u_D]\left(\overline{d^C}_B\, \overline{u^C}_A\, u_D\, e_C\right) \nonumber \\
& + & k[u^C_A,d^C_B,\nu_C,d_D]\left(\overline{d^C}_B\, \overline{u^C}_A\, d_D\, \nu_C\right)\,+\, {\rm h.c.}\,,\eeqa
with
\beqa \label{coff_gauge}
k[u_A,d_B,e^C_C,u^C_D] &=& \frac{g_{10}^2}{M_X^2}\left[\left(U_{e^C}^\dagger U_d\right)_{CB} \left(U_{u^C}^\dagger U_{u}\right)_{DA} + \left(U_{e^C}^\dagger U_u\right)_{CA} \left(U_{u^C}^\dagger U_d \right)_{DB} \right]\,,\nonumber \\
k[u^C_A,d^C_B,e_C,u_D] &=& -\frac{g_{10}^2}{M_X^2} \left(U_{d^C}^\dagger U_e\right)_{BC} \left(U_{u^C}^\dagger U_{u}\right)_{AD}-\frac{g_{10}^2}{M_Y^2} \left(U_{d^C}^\dagger U_u\right)_{BD} \left(U_{u^C}^\dagger U_e \right)_{AC}\,,\nonumber \\
k[u^C_A,d^C_B,\nu_C,d_D] &=& \frac{g_{10}^2}{M_X^2} \left(U_{d^C}^\dagger U_\nu\right)_{BC} \left(U_{u^C}^\dagger U_d\right)_{AD} + \frac{g_{10}^2}{M_Y^2} \left(U_{d^C}^\dagger U_d\right)_{BD} \left(U_{u^C}^\dagger U_\nu \right)_{AC}\,.\eeqa

Decay widths of proton in various channels then can be obtained by using $y[f_1,f_2,f_3,f_4]=k[f_1,f_2,f_3,f_4]$ and $y^\prime[f_1,f_2,f_3,f_4]=0$ in the expressions listed in Eq. (\ref{decay_width}). Flavour effects arising through various unitary matrices are evaluated from the best-fit solutions determined in \cite{Mummidi:2021anm} for the minimal non-supersymmetric $SO(10)$ model with Peccie-Quinn symmetry. The results obtained for the best fit solution and for different hierarchy among the $X$ and $Y$ gauge bosons are given in Table \ref{tab:gauge_res}. We also give the spectrum of branching ratios for several solutions with acceptable $\chi^2$ in Fig. \ref{fig3}. 
%%%%%%%%%%%%%
\begin{table}[h]
\begin{center}
\begin{tabular}{lccc} 
\hline
\hline
Branching ratio [\%]& ~~~$M_X \ll M_{Y}$~~~  & ~~~$M_X \gg M_{Y}$~~~ & ~~~$M_X = M_{Y}$~~~\\
\hline
${\rm BR}[p\to e^+ \pi^0]$ & $63$ & $33$ & $47$\\
${\rm BR}[p\to \mu^+ \pi^0]$  & $1$ & $1$ & $< 1$\\
${\rm BR}[p\to \bar{\nu} \pi^+]$  & $26$ & $55$ & $46$\\
${\rm BR}[p\to e^+ K^0]$ & $< 1$ & $< 1$ & $< 1$\\
${\rm BR}[p\to \mu^+ K^0 ]$  & $9$ & $<1$ & $4$\\
${\rm BR}[p\to \bar{\nu} K^+]$ & $3$ & $15$ & $1$\\
${\rm BR}[p\to e^+ \eta]$ & $<1$ & $<1$ & $<1$\\
${\rm BR}[p\to \mu^+ \eta]$  & $<1$ &  $<1$ & $< 1$\\
\hline
\end{tabular}
\end{center}
\caption{Proton decay branching fractions estimated for the best fit solution for various hierarchies among the masses of $X$ and $Y$ gauge bosons in case of the minimal renormalizable non-supersymmetric $SO(10)$ model.}
\label{tab:gauge_res}
\end{table}
%%%%%%%%%%%%%%%%%%%
%%%%%%%%%%%%%%%
\begin{figure}[t]
\centering
\subfigure{\includegraphics[width=0.325\textwidth]{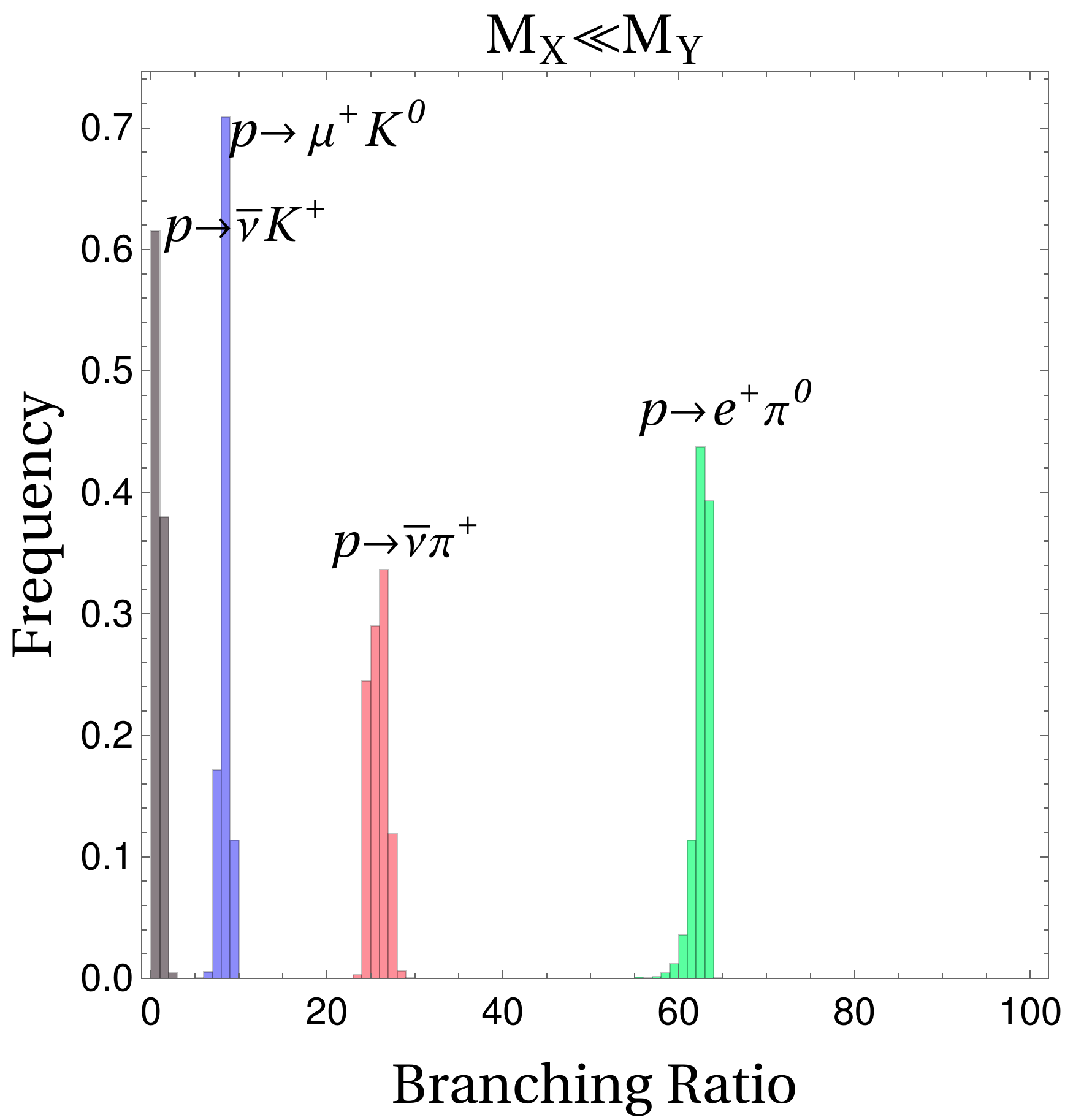}}
\subfigure{\includegraphics[width=0.325\textwidth]{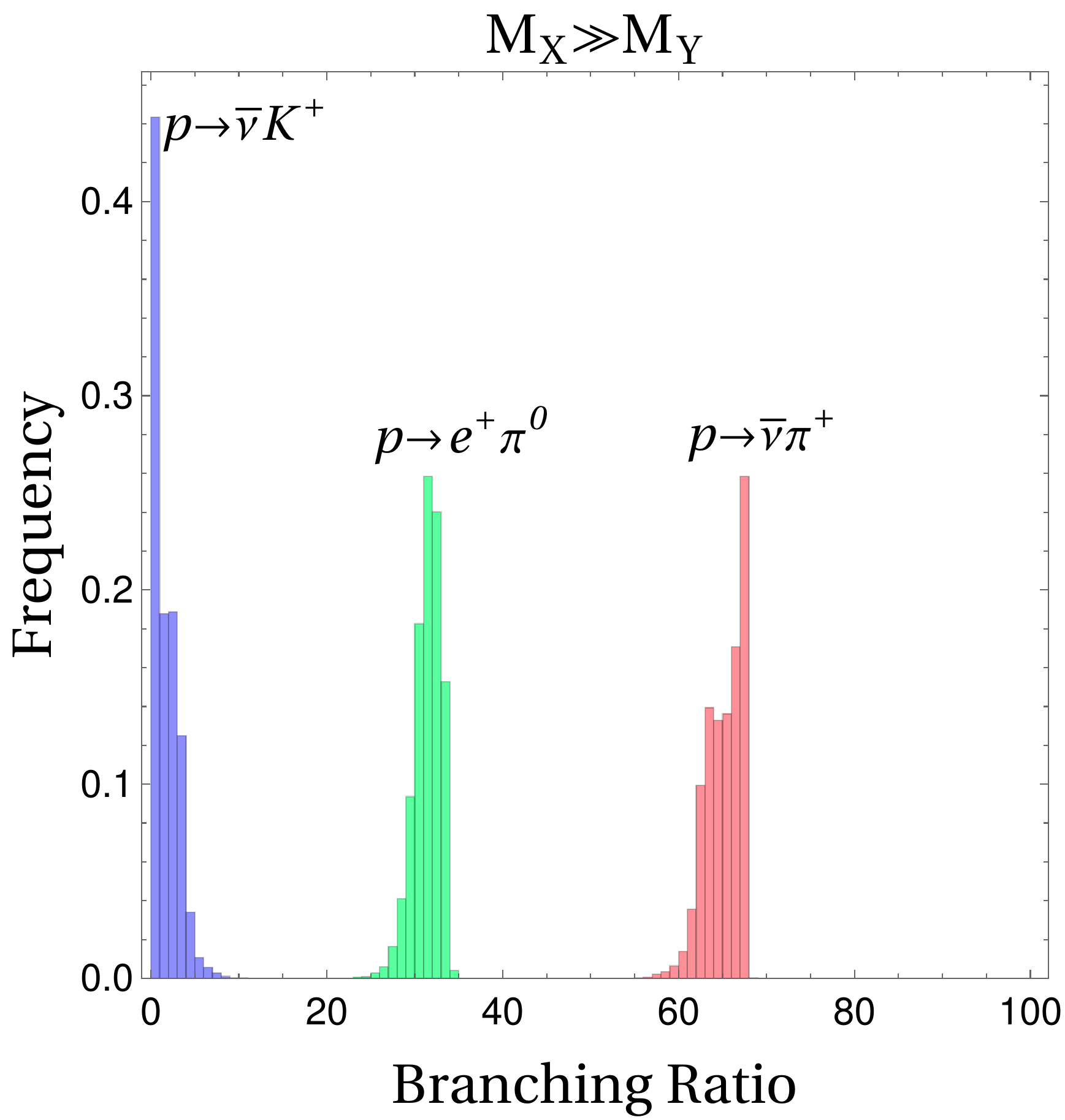}}
\subfigure{\includegraphics[width=0.325\textwidth]{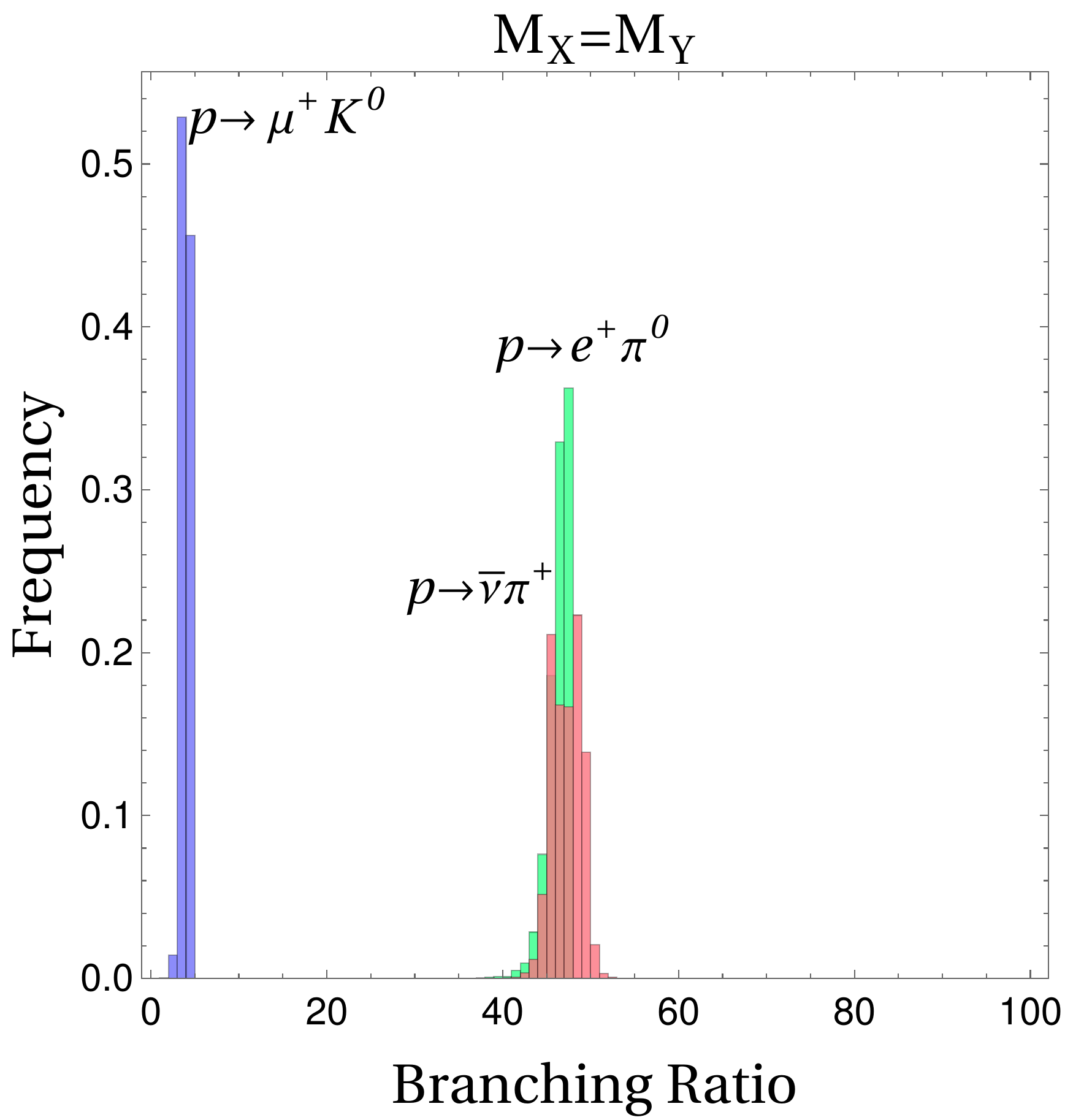}}
\caption{Gauge boson mediated proton decay spectrum for several solutions obtained in case of the minimal renormalizable non-supersymmetric $SO(10)$ model.}
\label{fig3}
\end{figure}
%%%%%%%%%%%%%%
It can be seen from Fig. \ref{fig3} that the results in Table \ref{tab:gauge_res} are robust predictions for the proton decays induced by the vector bosons. Proton decay branching ratios are qualitatively very different from those obtained by scalar mediation.

\newpage

%\bibliography{references.bib}
\bibliography{references}

\end{document}